\documentclass[preprint,11pt]{elsarticle}

\usepackage{amssymb}
\usepackage{amsthm}

\usepackage{comment}
\usepackage{xcolor}
\usepackage[edges]{forest}
\usepackage{booktabs}
\usepackage{listings}
\usepackage[margin=1in]{geometry}
\usepackage{array}

\definecolor{gray}{rgb}{0.75,0.75,0.75}
\definecolor{darkgray}{rgb}{0.85,0.85,0.85}

\usepackage{longtable}
\usepackage{lineno}

\newcounter{bla}

\usepackage{mathtools}
\usepackage{amsmath}
\usepackage{hyperref}
\usepackage{float}
\usepackage{subcaption}
\usepackage{graphicx}
\usepackage{textcomp,mathcomp}
\usepackage{bm}
\usepackage{multirow}
\usepackage{natbib}
\usepackage{bookmark}
\usepackage{cases}
\usepackage{tabularx}

\usepackage{pdflscape} 
\usepackage{tikz}
\usepackage{tikz-qtree}

\usepackage[T1]{fontenc} 
\usepackage{url}
\usepackage{breakurl} 

\journal{Computer Physics Communications}

\begin{document}

\begin{frontmatter}



\title{M2C: An Open-Source Software for Multiphysics Simulation of Compressible Multi-Material Flows and Fluid–Structure Interactions}


\author[a,b]{Xuning Zhao}
\author[a,c]{Wentao Ma}
\author[a]{Shafquat Islam}
\author[a]{Aditya Narkhede}
\author[a]{Kevin Wang\corref{author}}

\cortext[author] {Corresponding author.\\\textit{E-mail address:} kevinwgy@vt.edu}
\address[a]{Kevin T. Crofton Department of Aerospace and Ocean Engineering, Virginia Tech, Blacksburg, VA 24061, USA}
\address[b]{School of Engineering, Brown University, Providence, RI 02912, USA}
\address[c]{Department of Biomedical Engineering, Duke University, Durham, NC 27708, USA}

\begin{abstract}
M2C (Multiphysics Modeling and Computation) is an open-source software for simulating multi-material fluid flows and fluid-structure interactions under extreme conditions, such as high pressures, high temperatures, shock waves, and large interface deformations. It employs a finite volume method to solve the compressible Navier-Stokes equations and supports a wide range of thermodynamic equations of state. M2C incorporates models of laser radiation and absorption, phase transition, and ionization, coupled with continuum dynamics. Multi-material interfaces are evolved using a level set method, while fluid-structure interfaces are tracked using an embedded boundary method. Advective fluxes across interfaces are computed using FIVER (FInite Volume method based on Exact multi-material Riemann problems). For two-way fluid-structure interaction, M2C is coupled with the open-source structural dynamics solver Aero-S using a partitioned procedure. The M2C code is written in C++ and parallelized with MPI for high-performance computing. The source package includes a set of example problems for demonstration and user training. Accuracy is verified through benchmark cases such as Riemann problems, interface evolution, single-bubble dynamics, and ionization response.   Several multiphysics applications are also presented, including laser-induced thermal cavitation, explosion and blast mitigation, and hypervelocity impact.


\noindent \textbf{PROGRAM SUMMARY}

\begin{small}
\noindent
{\em Program Title:} M2C (Multiphysics Modeling and Computation)    \\
{\em CPC Library link to program files:} (to be added by Technical Editor) \\
{\em Developer's repository link:}  \url{https://github.com/kevinwgy/m2c} \\
{\em Licensing provisions:} GNU General Public License 3 \\
{\em Programming language:} C++  \\
{\em Nature of problem:}\\ This work addresses the analysis of multi-material fluid flow and fluid-structure interaction problems under conditions involving high pressure, high velocity, high temperature, or a combination of them. In such problems, material compressibility and thermodynamics play a significant role, and the system may exhibit shock waves, large structural deformations, and large deformation of fluid material subdomains. Unlike conventional fluid dynamics problems, the boundaries of the fluid domain and material subdomains are time-dependent, unknown in advance, and must be determined as part of the analysis. Across material interfaces, some state variables (e.g., density) may exhibit jumps of several orders of magnitude, while others (e.g., normal velocity) remain continuous. Some problems may also involve strong external energy sources, such as lasers, and energy deposition is coupled with fluid dynamics.  In certain cases, additional physical processes --- such as phase transition (e.g., vaporization) and ionization --- may arise and must be incorporated in the analysis. Example problems presented in this work include laser-induced cavitation, underwater explosion, blast mitigation, and hypervelocity projectile impact. More broadly, this type of problems are relevant to many engineering and biomedical applications, where understanding continuum mechanics and material behaviors under extreme conditions is essential.\\
{\em Solution method}:\\ The core of M2C is a three-dimensional finite volume solver for compressible flow dynamics. It is designed to support arbitrary convex equations of state in a modular fashion. Several models are currently implemented, including Noble-Abel stiffened gas, Jones-Wilkins-Lee (JWL), Mie-Gr\"uneisen, Tillotson, and an example of ANEOS (ANalytic Equation Of State). These models allow M2C to analyze a wide range of materials. M2C uses the level set method to track massless interfaces between fluid materials, providing sharp interface representation and supporting topological changes such as merging and separation. For fluid-structure interfaces, it employs an embedded boundary method that simplifies mesh generation and accommodates large structural deformation. Across material interfaces, M2C uses the FIVER (FInite Volume method with Exact multi-material Riemann problems) method to compute the advective fluxes, which is robust in the presence of large jumps in state variables.   M2C implements a partitioned procedure that enables two-way coupling with an external structural dynamics solver, exchanging data at each time step. It has been coupled with the open-source Aero-S solver for fluid-structure coupled analysis. Additional features include a latent heat reservoir method for vaporization and a multi-species, non-ideal Saha equation solver for material ionization. The M2C code is parallelized with MPI for high-performance computing and designed with modularity and object-oriented principles for ease of extension and reuse.\\
{\em Supplementary material:} \\ The M2C package includes a suite of test cases that illustrate the software's capabilities. These examples can also serve as templates for setting up new simulations.
%

\end{small}

\end{abstract}
\end{frontmatter}

\section{Introduction}
\label{sec:intro}
M2C stands for Multiphysics Modeling and Computation. It is an open-source scientific computing software for simulating the dynamics of continua in the Eulerian reference frame. M2C focuses on capturing the interactions of multiple physical processes --- such as mechanical and thermal --- across interfaces between different types of materials, including gases, liquids, and solids. It is particularly well-suited for analyzing multi-material fluid flows and fluid-structure interactions involving high pressures, shock waves, high temperatures, and phase transitions. The capabilities of M2C have been demonstrated through its application to a range of problems, including cavitation, underwater explosion, blast mitigation, and hypervelocity impact~\cite{zhao2023simulating,zhao2024vapour,MA2023112474,islam2023fluid,islam2023plasma,narkhede2025fluid,ma2024data,zhao2023numerical,islam2025ionization}. These examples illustrate the software’s versatility and potential for addressing  important problems in engineering and biomedical research, where understanding continuum mechanics and material behaviors under extreme conditions is essential.

The core of M2C is a three-dimensional (3D) finite volume compressible flow solver. It is equipped with various state reconstruction schemes, slope limiters, and numerical flux functions, including local Lax-Friedrichs (also known as Rusanov flux~\cite{rusanov1970difference}), Roe-Pike~\cite{hu2009hllc}, and HLLC~\cite{toro2013riemann}.  Unlike many conventional solvers that support only simple equations of state (EOS) --- such as perfect and stiffened gases --- M2C is designed to support arbitrary convex EOS in a modular fashion. Currently implemented EOS include Noble-Abel stiffened gas, Jones-Wilkins-Lee (JWL), Mie-Gr\"uneisen, Tillotson, and an example of ANEOS (ANalytical EOS)~\cite{NobelAbel2016,menikoff2015jwl,Robinson2019mie,brundage2013implementation,sanchez2021inelastic}. These models allow M2C to simulate a wide variety of materials. Another feature of M2C is its ability to handle multiple material subdomains separated by sharp interfaces, with a separate EOS assigned within each subdomain. Across material interfaces, advective fluxes are computed using the FIVER (FInite Volume method based on Exact two-material Riemann problems) method, which enforces continuity of pressure and normal velocity~\cite{farhat2012fiver,cao2021shock,wang2017multiphase, ma2022computational}. 

M2C captures the dynamics of material interfaces using level set and embedded boundary methods~\cite{ma2022computational,wang2012computational}. For massless interfaces transported by the velocity field (e.g.,~fluid-fluid interfaces), their motion and deformation are captured by solving level set equations within either the entire computational domain or a narrow band of elements around the interface. Interfaces of other types, such as fluid-structure interfaces, can be represented as triangulated surfaces embedded within the computational domain. Their dynamics can be either specified by the user through an input file or computed by an external solver that operates concurrently with M2C, exchanging data in real-time. Currently, M2C is coupled using a partitioned procedure with Aero-S~\cite{aeros}, an open-source, parallelized nonlinear finite element solver for structural and solid mechanics. This embedded boundary fluid-structure coupling framework is well-suited for simulating fluid-structure interactions involving large structural deformation and topological changes.

M2C includes a module for simulating laser radiation and absorption by solving a specialized form of the radiative transfer equation (RTE).  This module is integrated with the compressible flow solver, enabling simulations that capture radiative heating effects from laser sources. M2C also includes a recently developed phase transition model, which allows it to simulate thermal cavitation processes~\cite{zhao2023simulating,zhao2024vapour}. Additionally, M2C provides a module for solving the non-ideal Saha equations to predict the onset of ionization~\cite{islam2023plasma,islam2025ionization}.

M2C is developed in C++ and parallelized using the message passing interface (MPI) for execution on distributed-memory CPU clusters. It utilizes several open-source scientific computing libraries, including Eigen~\cite{guennebaud2010eigen}, Boost~\cite{schaling2011boost}, and PETSc~\cite{balay2020petsc}. In many cases, running M2C only requires the user to create a text input file, which specifies the physical models, computational domain and mesh resolution, boundary and initial conditions, numerical methods, and associated parameters. M2C offers flexible output options: solution fields can be saved over the entire domain, on user-defined planes or lines, or at specified sensor locations. Full-domain results are written in VTK format, compatible with many open-source and commercial visualization tools.  In addition, M2C allows users to define custom initial and boundary conditions through user-written functions with access to mesh and state variable data, which can be compiled separately and linked dynamically to M2C at runtime.

M2C shares some features with other open-source simulation packages for compressible multi-material flows, while also offering distinct capabilities. For example, the Aero-F solver~\cite{aerof} also implements the embedded boundary, level set, and FIVER methods for multi-material fluid and fluid-structure simulations, but it operates on unstructured tetrahedral meshes. In contrast, M2C uses Cartesian meshes, and offers better efficiency for problems with cylindrical or spherical symmetry. Laser radiation, phase transition, and ionization modeling capabilities are currently not available in Aero-F. Additionally, M2C employs a generalized, accelerated bimaterial Riemann problem solver for computing interfacial fluxes~\cite{MA2023112474}. M2C also shares capabilities with the Multicomponent Flow Code (MFC)~\cite{bryngelson2021mfc}, such as simulating compressible multi-material flows with shock waves and complex interface dynamics. A key distinction lies in interface treatment. MFC uses a diffusive interface capturing method, while M2C employs sharp interface tracking with level set and embedded boundary methods. Although both solvers have phase transition models, their approaches differ substantially due to different interface representation strategies. Furthermore, OpenFOAM~\cite{weller1998tensorial}, a general-purpose computational fluid dynamics (CFD) platform, also supports multi-component compressible flow simulations using the volume-of-fluid (VOF) method, which yields diffused interfaces. It does not offer bimaterial Riemann solvers and flexible EOS support, often requiring significant user customization for extension. In comparison, M2C overcomes these limitations through native support for arbitrary convex EOS, bimaterial Riemann solvers, and sharp interface methods, along with integrated models for laser-fluid coupling, ionization, and fluid-structure coupling.

The remainder of this paper is organized as follows. Section~\ref{sec:overview} provides an overview of the M2C solver, including its organization, key features, and structure. Section~\ref{sec:equation} describes the physical models employed in M2C, including governing equations, general EOS, interface conditions, and models of laser radiation, phase transition, and ionization. Section~\ref{sec:numerical_method} outlines the numerical methods used to solve these equations. Section~\ref{sec:VandV} presents verification and validation test cases. Section~\ref{sec:example} showcases M2C’s capabilities through illustrative examples, including non-spherical bubble formation induced by laser radiation, plasma generation in fluids from hypervelocity impacts, and large deformations of a lightweight explosion-containment chamber. Finally, Section~\ref{sec:conclusion} provides a brief summary.

\section{Overview and features}
\label{sec:overview}
\subsection{Installation and usage}
\label{sec:install}

M2C is freely available at \url{https://github.com/kevinwgy/m2c}. Before compiling M2C, several open-source libraries need to be installed, including Eigen~\cite{guennebaud2010eigen}, Boost~\cite{schaling2011boost}, and PETSc~\cite{balay2020petsc}. Eigen and Boost are header-only, which simplifies the setup. An MPI library is required for parallel execution, and CMake is used to generate the build files.

In general, M2C can be compiled in a Linux environment using the following commands:
\begin{lstlisting}[language=bash]
  $ cd [main directory of M2C]
  $ cmake .              
  $ make -j [number of processes]
\end{lstlisting}
If CMake does not automatically locate all the required libraries and dependencies, the \texttt{CMakeLists.txt} file may need to be customized to help CMake find them.

Upon successful compilation, the executable \texttt{m2c} will be generated. Users can verify the installation by running some example cases provided in the \texttt{Tests} directory. A typical command to launch a simulation is:
\begin{lstlisting}[language=bash]
$ mpiexec -n [number of MPI processes] [M2C executable] input.st
\end{lstlisting}
where \texttt{input.st} is the input text file specifying the simulation setup.

\begin{figure}
    \centering
    \includegraphics[width=1.0\linewidth]{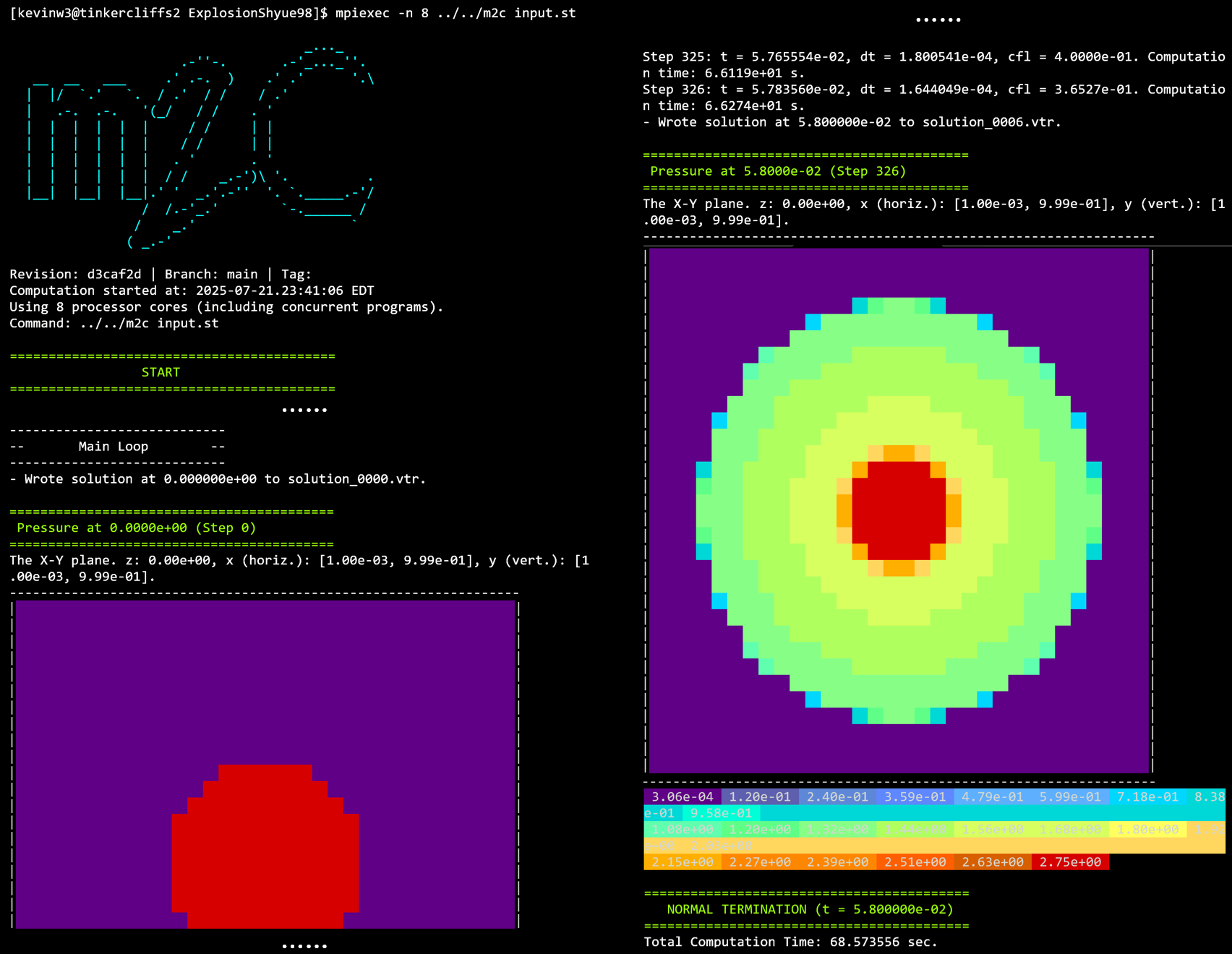}
    \caption{Screenshots from a test run.}
    \label{fig:test_run}
\end{figure}

For instance, to run the example in \texttt{Tests/ExplosionShyue98} using $8$ parallel processes (typically corresponding to $8$ CPU cores), the user can navigate to that directory and execute the following command:
\begin{lstlisting}[language=bash]
$ mpiexec -n 8 ../../../m2c input.st
\end{lstlisting}

Figure~\ref{fig:test_run} shows a few snapshots of the screen output. The displayed information includes simulation metadata (such as code version, start time, number of parallel processes, and execution command), simulation setup, and time-stepping progress. In addition, M2C supports terminal-based visualization of some important field variables---such as pressure, velocity magnitude, and material ID---which can be useful during early-stage trial runs for quick diagnostics.

\subsection{Features}
\label{sec:features}

M2C provides a range of capabilities for simulating multi-material and multiphysics problems. Some important features are outlined below.

\begin{enumerate}[1.]
    \item Computational architecture
    \begin{itemize}
        \item Supports 1D, 1D-spherical, 2D, 2D-cylindrical, and 3D spatial domains.
        \item Built-in mesh generator: uniform and non-uniform Cartesian grids.
        \item Domain decomposition for distributed parallel computing.
    \end{itemize}
    \item Multi-material flow modeling
    \begin{itemize}
        \item Solves compressible flow equations for both single-phase and multi-material fluids with sharp interfaces.
        \item Level set-based interface tracking with multiple reinitialization schemes.
        \item Supports arbitrary convex Equation of State (EOS) models and allows adding new EOS without modifying the core solver.  
        \item Models phase transitions (e.g., vaporization) using a latent heat reservoir method. 
    \item Additional physical effects:
    \begin{itemize}
        \item Viscosity
        \item Heat conduction
        \item Volumetric forces (e.g., gravity)
    \end{itemize}
    \end{itemize}
    \item Numerical schemes
    \begin{itemize}
        \item Second-order accurate finite-volume discretization using the MUSCL scheme.
        \item Supports reconstruction in primitive, conservative, or characteristic variables.
        \item Various slope limiters (MC, Van Albada, Modified Van Albada) for TVD stability.
        \item Multiple numerical flux functions: Local Lax-Friedrichs, Roe-Pike, and HLLC.
        \item FIVER (Finite Volume Method with Exact Riemann Solvers) for resolving fluxes across material interfaces.
        \item Efficient bimaterial Riemann problem solver accelerating computation without compromising solution accuracy and solver robustness.
        \item High-order time integration with 2nd- and 3rd-order Runge-Kutta TVD schemes and Forward Euler for debugging.
    \end{itemize}
    \item Multiphysics extensions
    \begin{itemize}
        \item Fluid-structure coupling using embedded boundary method and compatible with external solid mechanics solvers (e.g., Aero-S).
        \item Laser propagation and energy deposition in fluids and capable of predicting the onset, expansion, and collapse of laser-induced bubbles.  
        \item Ionization and plasma expansion for gases by solving ideal or non-ideal Saha equations.
    \end{itemize}   
\end{enumerate}

\subsection{Code structure and workflow}
\label{sec:code_structure}

The M2C package currently consists of a few hundred files and several sub-directories. To provide a clearer understanding of the code organization, we categories them into several components as outlined in Table~\ref{tab:files_overview}. These components represent key functionalities, including the finite volume solver for compressible flow, level set method for interface tracking, multi-material modeling, fluid-structure interaction, laser energy deposition, ionization, and output processing. The specific files and directories included in each component are detailed in \ref{app:files_org}. While these components are not actual directories within the code, they serve as a conceptual framework to clarify the relationships between different files.

\begin{table}
    \centering
    \small
    \renewcommand{\arraystretch}{1.2}
    \begin{tabular}{p{2.3cm}|p{3.8cm}|p{9cm}}
        \toprule
        \multicolumn{2}{l|}{\textbf{Component}} & \textbf{Description} \\
        \hline
        \multicolumn{2}{l|}{\textbf{Input}} & Manages input data, user-defined states, and prescribed motion \\
        \hline
        \multicolumn{2}{l|}{\textbf{Mesh generation}} & Handles mesh generation, partition, and storage \\
        \hline
        \multirow{5}{{2.3cm}}{\textbf{Finite-volume solver}} & Equations of State (EOS) & Includes a base class and multiple specific models inheriting the base class\\
        \cline{2-3}
        & Multi-material fluids  &  Implements the main components, including time integration, multiphase modeling, and spatial initialization\\
        \cline{2-3}
        & Exact Riemann solver & Solves the two-material Riemann problem exactly, with acceleration methods in~\cite{MA2023112474} \\
        \cline{2-3}
        & Flux computation & Handles numerical flux schemes and reconstruction methods\\
        \cline{2-3}
        & Additional physics & Includes gravity, heat diffusion, and viscosity models \\
        \hline
        \multicolumn{2}{l|}{\textbf{Level set}} & Tracks material interfaces using a level set method with reinitialization\\
        \hline
        \multicolumn{2}{l|}{\textbf{Fluid-Structure Interaction (FSI)}} & Supports embedded boundary methods for FSI\\
        \hline
        \multicolumn{2}{l|}{\textbf{Ionization}} & Ionization models and Saha equation solvers\\
        \hline
        \multicolumn{2}{l|}{\textbf{Laser}}  & Simulates laser energy absorption, deposition, and transport \\
        \hline
        \multicolumn{2}{l|}{\textbf{Phase change}} & Models phase transitions and latent heat effects\\
        \hline
        \multicolumn{2}{l|}{\textbf{Output}} & Controls simulation output, post-processing, and visualization \\
        \hline
        \multirow{4}{*}{\textbf{Utilities}} & MPI communication & Provides MPI-based parallel communication tools \\
        \cline{2-3}
        & External solver coupling & Interfaces with external solvers for multi-solver setups \\
        \cline{2-3}
        & Computational tools & Provides general computational and numerical tools \\
        \cline{2-3}
        & Compilation & Includes build configuration and licensing files \\
        \hline
        \multicolumn{2}{l|}{\textbf{Tests}} & A collection of test cases, each containing input files, log files, and representative results. These can serve as references for setting up new simulations \\
        \bottomrule
    \end{tabular}
    \caption{Classification of M2C files and directories based on functionality. Details of the specific files and directories in each component are provided in Table~\ref{tab:files}}
    \label{tab:files_overview}
\end{table}

\begin{figure}
    \centering
    \includegraphics[width=0.8\linewidth]{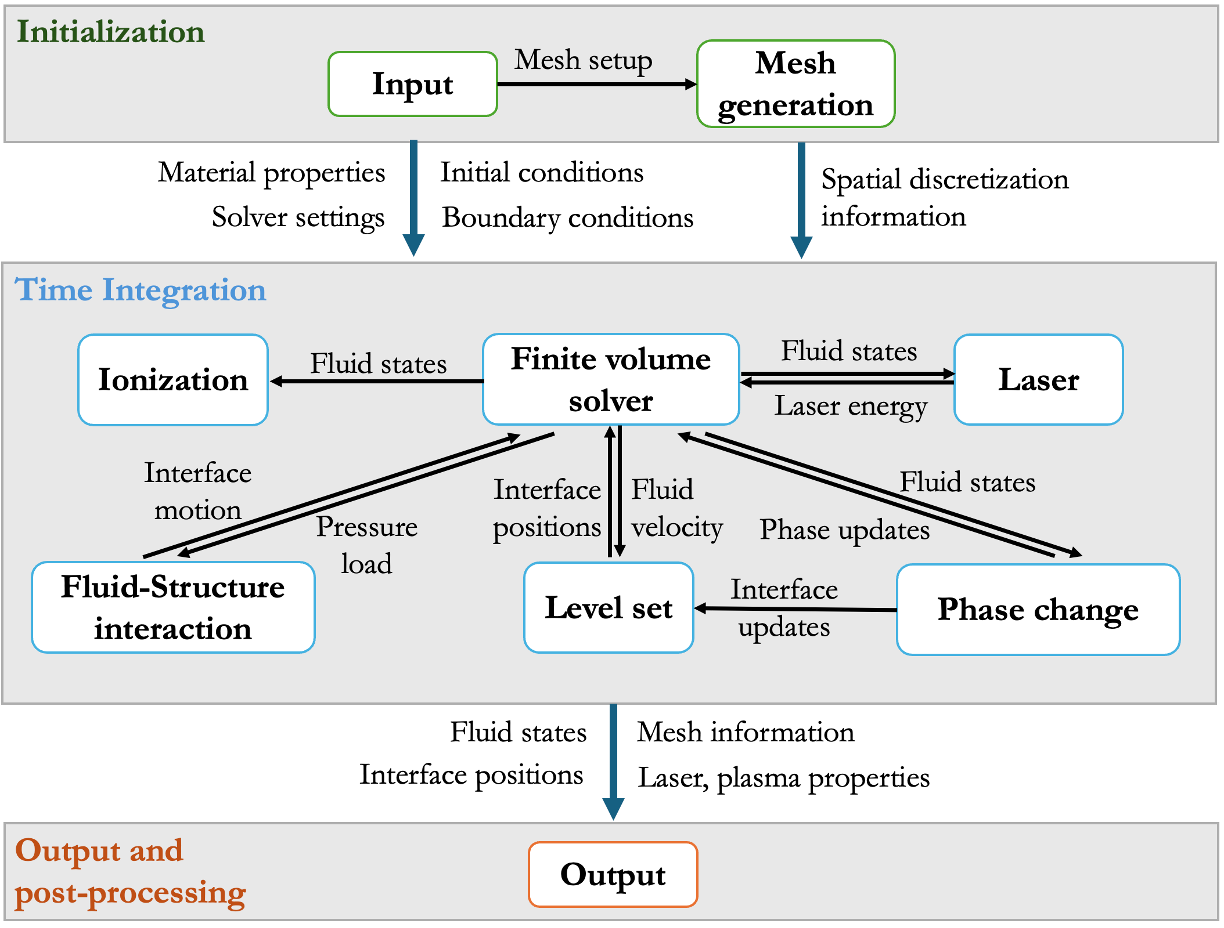}
    \caption{Code structure and data exchange in M2C.}
    \label{fig:code_structure}
\end{figure}

The execution flow of M2C is structured into three main stages, initialization, time integration, and output and post-processing. Figure~\ref{fig:code_structure} provides a visual representation of these stages, showing the flow of information between key components. Specifically, the simulation begins with input processing, where material properties, solver settings, initial conditions, and boundary conditions are defined. The mesh generation module constructs the computational domain based on input specifications. During time integration, the finite volume flow solver interacts with various physical models, including thermodynamic equations of state, ionization, phase change, and laser energy absorption, to update the state variables. The level set method tracks massless material interfaces, while the fluid-structure interaction module exchanges fluid-induced force loads and interface motion between M2C and a structural solver. Finally, the output and post-processing stage extracts and stores relevant simulation results such as flow fields, interface positions, and mesh data.

The details of input and output files are shown in Section~\ref{sec:code_I/O}. In the time integration stage, M2C follows a structured sequence of operations during each time step. An outline of the time integration procedure within one time step is provided below. The solver first computes residuals for fluid governing equations and updating the fluid states, followed by level set updates for interface tracking. If phase change module is activated, it modifies the fluid state accordingly. Additional physics, such as laser heating and ionization, are integrated where applicable.

\noindent\begin{tabularx}{\textwidth}{cll}
\bf{Input:} & \multicolumn{2}{l} {Numerical solution at the previous time step}\\
(1) & \multicolumn{2}{l} {Compute the residual of the Navier-Stokes equations}\\
 & (1.1) Compute the advective fluxes. & \\
 & (1.2) Compute the diffusive heat fluxes & (if heat diffusion is enabled). \\
 & (1.3) Compute the viscous terms & (if viscosity modeling is active). \\
 & (1.4) Compute the radiative heat source & (if laser is enabled). \\
 & (1.5) Compute the body force source & (if gravity is included). \\
(2) & \multicolumn{2}{l} {Advance the fluid state by one time step. Apply boundary conditions.}\\
(3) & \multicolumn{2}{l} {Compute the residual of the level set equation.}\\
(4) & \multicolumn{2}{l} {Advance the level set function by one time step. Apply boundary conditions.}\\
(5) & \multicolumn{2}{l} {Reinitialize the level set equation, if reinitialization is scheduled based on the }\\
 & \multicolumn{2}{l} {prescribed frequency.}\\
(6) & \multicolumn{2}{l} {Update phase using level set value. Update fluid state at nodes that changed phase }\\
& \multicolumn{2}{l}{due to interface motion.}\\
(7) & Check for phase transition & (if phase change modeling is activated).\\
&  \multicolumn{2}{l} {Update fluid state at nodes that have undergone phase transition.}\\
(8) & \multicolumn{2}{l} {If phase transition occurred, reinitialize the level set equation.}\\
(9) & Solve the laser radiation equation & (if laser modeling is enabled).\\
(10) & Solve the Saha equation & (if ionization effects are considered).\\
(11) & \multicolumn{2}{l} {Exchange data with the solid solver or update the interface position from input.}\\
& & (if the embedded boundary method is used). \\
\bf{Output:} & \multicolumn{2}{l} {Numerical solution at this step.}
\end{tabularx}

\subsection{Input and output}
\label{sec:code_I/O}

In M2C, the simulation setup is specified through a modular text file, which organizes physical and numerical options and parameters into multiple sections. In the example problems provided in the \texttt{Tests} directory, this file is always named \texttt{input.st}. Table~\ref{tab:input_overview} provides an overview of the main input sections.

\begin{table}
    \centering
    \small
    \renewcommand{\arraystretch}{1.2} 
    \begin{tabular}{p{4.6cm}|p{10.6cm}}
        \hline
        \textbf{Input file class} & \textbf{Description} \\
        \hline
        \texttt{Mesh} & Defines computational domain type, boundaries, corresponding boundary conditions, and mesh generation parameters. \\
         \hline
         \texttt{ConcurrentPrograms} & Handles coupling with other solvers and the coupling algorithm.\\
         \hline
         \texttt{EmbeddedBoundaryMethod} & Specifies embedded surfaces within computational domain, including geometry, mesh, boundary conditions, and numerical parameters. \\
        \hline
         \texttt{Equations} & Defines material properties, including EOS, viscosity, and heat diffusion. Also includes parameters for phase transition.\\
         \hline 
         \texttt{InitialCondition} & Specifies initial conditions such as material ID, density, velocity, and pressure, with support for multiple regions. \\
         \hline
         \texttt{BoundaryCondition} & Specifies boundary conditions based on boundary types, such as inlet, farfield, and wall. \\
         \hline
         \texttt{ExactRiemannSolution} & Defines numerical parameters required for solving two-material Riemann problems exactly at material interfaces.\\
         \hline
         \texttt{Space} & Specifies numerical parameters for solving Navier-Stokes equations and level set equations, such as flux calculation method.\\
         \hline
         \texttt{Time} & Defines time integration settings, including time-stepping schemes and simulation duration. \\
         \hline
         \texttt{MultiPhase} & Handles numerical schemes for treating material interfaces in multiphase/multi-material simulations. \\
         \hline
         \texttt{Laser} &  Defines laser-related properties such as source position, power profile, and material absorption coefficients. \\
         \hline
         \texttt{Ionization} & Specifies ionization-related material properties and numerical parameters for solving the Saha equation. \\
         \hline
         \texttt{SpecialTools} & Enables optional computational utilities, including dynamic load calculation and EOS tabulation.\\
         \hline
         \texttt{Output} & Controls the saving of simulation results, including output frequency, file paths, and selected physical variables. Also manages the setup and output of post-processing tools.\\
        \hline
        \texttt{TerminalVisualization} & Configures real-time terminal-based visualization.\\ 
         \hline
    \end{tabular}
    \caption{Overview of Input File Structure in M2C.}
    \label{tab:input_overview}
\end{table}

A complete list of input parameters is available in \texttt{IoData.h/cpp}, where each parameter’s default value is set in \texttt{IoData.cpp}. If a parameter is omitted in the input file, the simulation automatically falls back to its default value. Some input data can also be specified via external files, such as user-prescribed initial conditions, mesh for embedded surfaces, and laser power time profiles. Example input files for various test cases are available in the \texttt{Tests} directory, which can be helpful for setting up new simulations. Not all input classes are required for every simulation. For example, the \texttt{laser} class is only needed when modeling laser radiation.

In \texttt{input.st}, the \texttt{Output} class defines the output control parameters, including file naming, output frequency, physical quantities, and post-processing functions. M2C could generate 3D solution fields in VTK format, with users specifying which parameters to output. By enabling \texttt{MeshInformation} and \texttt{MeshPartition} parameters, the code can output detailed mesh information. M2C also allows solution data to be extracted at user-defined points, lines, and planes by activating \texttt{Probes}, \texttt{LinePlot}, and \texttt{CutPlane}, respectively. Additionally, the \texttt{EnergyIntegration} feature enables energy evolution analysis over user-defined regions. 

Real-time visualization is supported through the \texttt{TerminalVisualizetion} class, which allows solutions to be displayed on an axis-aligned plane directly on the screen. M2C also provides built-in functions for exporting data, including \texttt{WriteToVTRFile}, which outputs any spatially distributed variable as a VTR file, and \texttt{WriteToMatlabFile}, which saves vectors and matrices in Matlab format. Further details on these functions can be found in \texttt{SpaceVariable3D.h/cpp} and \texttt{LinearOperator.h/cpp}.

\section{Computational domain and model equations}
\label{sec:equation}

\subsection{Material subdomains}
\label{sec:subdomains}

M2C adopts the Eulerian reference frame and the Cartesian coordinate system to describe kinematics and dynamics. The overall computational domain, denoted by $\Omega$, is always an axis-aligned rectangular box, fixed in time. Within $\Omega$, there may be multiple subdomains, each representing a specific material type. These subdomains may have arbitrary shapes, and their boundaries can evolve over time. A subdomain does not have to be a connected subset of $\Omega$. For example, if there are two disjoint bubbles that have the same gas content, they can be represented by a single material subdomain. In general, different model equations or material properties are assigned to each subdomain. In some cases, $\Omega$ may include ``inactive'' subdomains, where no equations are solved by M2C. A subdomain can also be empty initially and only forms during the simulation, such as in problems involving phase transitions.

\begin{figure}[H]
    \centering
    \includegraphics[width=0.6\linewidth]{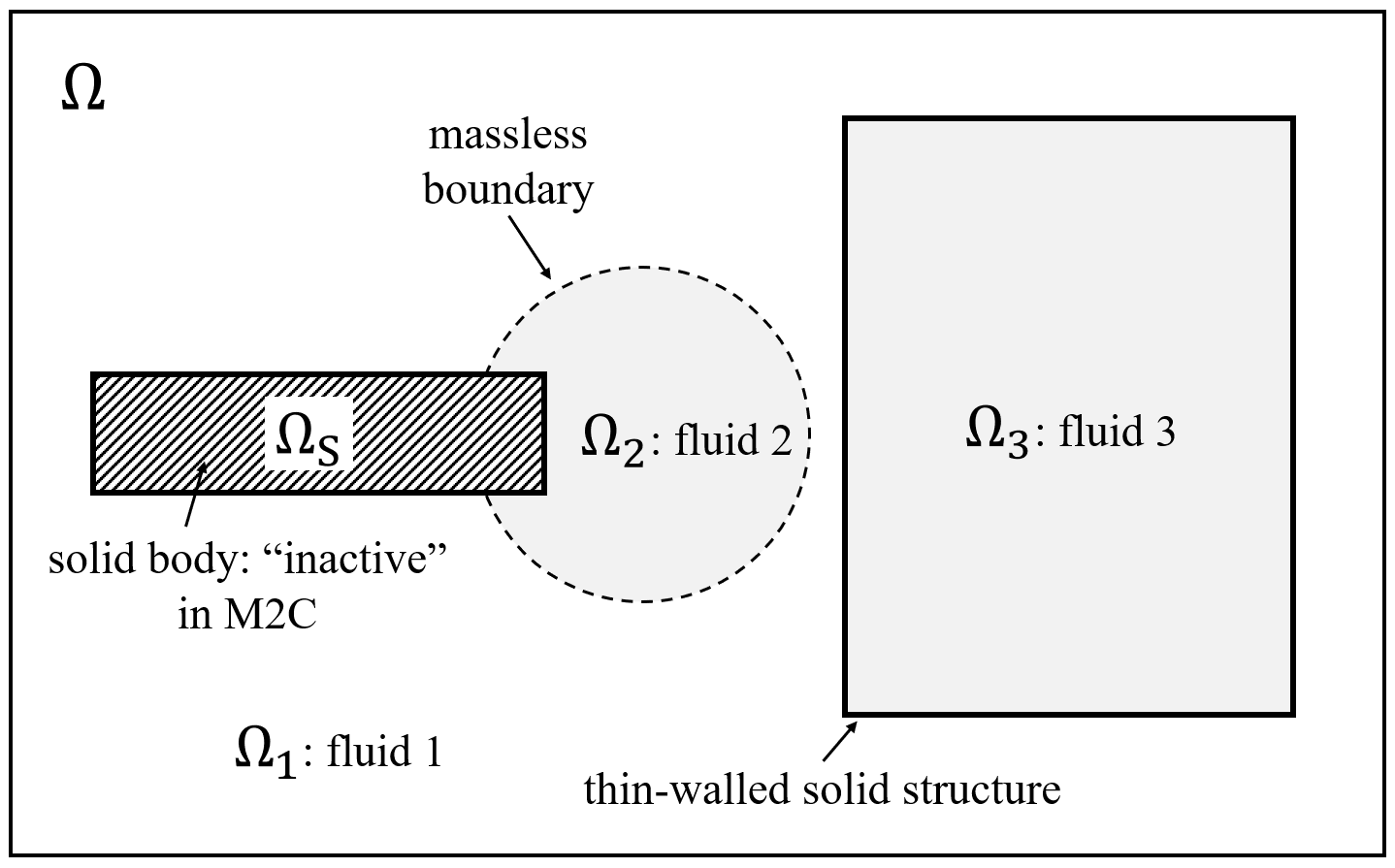}
    \caption{Computational domain for a multi-material fluid-structure interaction problem.}
    \label{fig:computational_domain}
\end{figure}

Figure~\ref{fig:computational_domain} illustrates an example problem featuring  a multi-material fluid flow interacting with a solid body and a thin-walled solid structure. In this case, $\Omega$ consists of an inactive region $\Omega_\text{S}$ corresponding to the space occupied by the solid body, and an active fluid region that is further divided into three non-overlapping subdomains, $\Omega_1$, $\Omega_2$, and $\Omega_3$, each representing   a different fluid material. Subdomain $\Omega_3$ is bounded by a thin-walled solid structure, while the interface between $\Omega_1$ and $\Omega_2$ is a massless internal boundary. The dynamics of subdomain boundaries $\partial\Omega_3$ and $\partial\Omega_\text{S}$ can be either prescribed by the user or computed using a structural dynamics solver coupled with M2C (e.g., Aero-S). The dynamics of the massless boundary $\partial\Omega_1\cap\partial\Omega_2$ is  determined by the local velocity field, and is tracked in M2C using the level set method.

M2C provides two methods for users to initialize subdomains:
\begin{itemize}
\item Explicit geometric specification: Subdomains can be defined using simple geometric primitives, including planes, spheres, parallelepipeds, spheroids, cones, circular cylinders, and circular cylinders with spherical endcaps. Examples: \texttt{Tests/BubbleInertCollapseCao21}, \texttt{Tests/RiemannProblems},  \texttt{Tests/HVImpactShafquatIslam/2D\_axisym\_HVI}.
\item Mesh-based definition: For more complex geometries, subdomains can be specified using a triangulated surface mesh that discretizes the subdomain boundary. This mesh can be provided directly through an input file (e.g., \texttt{Tests/LaserCavitationZhao24}) or, in fluid-structure interaction (FSI) simulations,  by a structural dynamics solver coupled to M2C (e.g., \texttt{Tests/UNDEXImplosion}).
\end{itemize}


\subsection{Governing equations}
\label{sec:NSEq}

In general, M2C solves the Navier-Stokes equations of the form
\begin{equation}
\frac{\partial \bm{W}}{\partial t} + \nabla \cdot \mathcal{F}(\bm{W}) = \nabla \cdot \mathcal{G}(\bm{W}) + \mathcal{S}(\bm{W}), \qquad t>0.
\label{eq:NSEquation}
\end{equation}
where $\bm{W}$ is the conservative state vector, $\mathcal{F}(\bm{W})$ represents the advective fluxes. $\mathcal{G}(\bm{W})$ includes the effects of viscosity, heat diffusion, and radiative heat transfer, based on user's specifications. The source term $\mathcal{S}$ accounts for body forces, such as gravity. In Cartesian coordinates, these terms can be expressed in a matrix and vector form as
\begin{equation}
\bm{W}=\begin{bmatrix} \rho \\ \rho \bm{V} \\ \rho E \end{bmatrix}, ~ \mathcal{F} = \begin{bmatrix} \rho \bm{V}^{T} \\  \rho \bm{V} \otimes \bm{V} + p \mathbb{I} \\ (\rho E + p) \bm{V}^T \end{bmatrix}, ~ \mathcal{G} = \begin{bmatrix} \bm{0}^T \\ \bm{\tau}\\   (\bm{\tau} \bm{V} + k\nabla T- \bm{q_r})^T \end{bmatrix}, ~ \mathcal{S} = \begin{bmatrix} 0 \\ \rho \bm{b} \\ \rho \bm{V}^T \bm{b} \end{bmatrix}.
\label{eq:NSEquation_terms}
\end{equation}
Here,  $\rho$, $p$, and $T$ denote density, pressure, and temperature, respectively. $\bm{V} = [u, v, w]^T$ is the velocity vector. $E$ is the total energy per unit mass, given by
\begin{equation}
E = e + \dfrac{1}{2} |\bm{V}|^2,
\end{equation}
where $e$ denotes internal energy per unit mass. $\mathbb{I}$ denotes the $3\times3$ identity matrix. $\bm{\tau}$ is the viscous stress tensor (a $3\times 3$ matrix). $k$ is the thermal conductivity coefficient, which varies based on the material. $\bm{q_r}$ is the radiative heat flux vector, and $\bm{b}$ denotes the body force vector per unit mass. 

\subsection{Thermodynamic equation of state}
\label{sec:EOS}

Within each material subdomain, a thermodynamic equation of state (EOS) is needed to close the system of governing equations. M2C is designed to accommodate any convex EOS expressed in the general form
\begin{align}
    p &= p (\rho, e),
    \label{eq:eos_all}\\
    T &= T(\rho, e).\label{eq:Temperature_law}
\end{align}

Implementing a new EOS in M2C is straightforward, as it only requires creating a derived class of \texttt{VarFcnBase}. In addition to \eqref{eq:eos_all} and \eqref{eq:Temperature_law}, the derived class should also define functions for evaluating $e(\rho,p)$, $\rho(p,e)$, $\displaystyle\dfrac{\partial p}{\partial \rho}(\rho, e)$,  $\displaystyle\dfrac{\partial p}{\partial e}(\rho, e)$, $e(\rho,T)$, and $e(\rho,h)$, where $h=e+p/\rho$ denotes enthalpy per unit mass. Some other material properties, such as speed of sound and bulk modulus, are computed using these functions internally.

Following this general framework, M2C currently includes several specific EOS implementations, including Noble-Abel stiffened gas (NASG)~\cite{NobelAbel2016}, Jones-Wilkins-Lee (JWL)~\cite{menikoff2015jwl}, Mie-Gr\"{u}neisen (MG)~\cite{Robinson2019mie}, Tillotson~\cite{brundage2013implementation}, and an example of ANEOS (ANalytic EOS) with Birch-Murnaghan equation and Debye model~\cite{sanchez2021inelastic}. These models span a wide range of materials and physical regimes, from compressible gases and condensed matter to detonation products and planetary materials. 

For example, the NASG EOS used in several numerical tests is defined by
\begin{equation}
p=(\gamma-1) \dfrac{\rho(e  - e_c)}{1  - \rho b}  - \gamma p_{c},
\label{eq:NASG_EOS}
\end{equation}
where $\gamma$, $e_c$, $b$, and $p_c$ are constant parameters. The temperature law (a slight generalization of the one proposed in~\cite{NobelAbel2016}) is given by
\begin{equation}
T(\rho,e) = \dfrac{1}{c_{v}} \Big(e-e_c -\big(\dfrac{1}{\rho}  -b\big)p_{c} - \Big(\dfrac{C}{1/\rho-b}\Big)^\gamma \dfrac{\big(1/\rho  -b\big)p_{c} }{\gamma-1}\Big),
\label{eq:NASG_Tlaw}
\end{equation}
where $c_v$ denotes the specific heat capacity at constant volume, and $C$ an additional constant parameter (set to $0$ in \cite{NobelAbel2016}). Other EOS available in M2C are presented in~\ref{app:eos}.

\subsection{Spherical and cylindrical symmetry}
\label{sec:symmetry_model}

For problems with spherical or cylindrical symmetry, M2C can solve the 3D governing equations~\eqref{eq:NSEquation} using a one-dimensional (1D) or two-dimensional (2D) computational domain, significantly reducing the computational cost. For simplicity, we consider the Euler equations here, i.e.,~\eqref{eq:NSEquation} with $\mathcal{G}(\bm{W}) = \bm{0}$ and $\mathcal{S}(\bm{W}) = \bm{0}$. The expressions of other terms (e.g., viscous and source terms) can also be derived through standard procedures and have been implemented in M2C; users may refer to the source code for details.

For problems with spherical symmetry, the solution is independent of the angular coordinates $\theta$ and $\phi$, depending only on the radial distance $r$ and time $t$. The state variables then reduce to
\begin{equation}
\Big(\rho(\bm{x},t),~\bm{V}(\bm{x},t),~E(\bm{x},t)\Big) = \Big(\rho(r,t),~u_r(r,t)\dfrac{\bm{x}}{r},~E(r,t)\Big),
\end{equation}
where $r = \sqrt{x^2 + y^2 + z^2}$. This leads to the following 1D governing equations
\begin{equation}
\dfrac{\partial}{\partial t}
\begin{bmatrix}
\rho\\ \rho u_r\\ \rho E
\end{bmatrix}
+ \dfrac{\partial}{\partial r}
\begin{bmatrix}
\rho u_r \\ \rho u_r^2 + p \\ (\rho E + p) u_r 
\end{bmatrix}
= -\dfrac{2}{r}\begin{bmatrix}
\rho u_r\\\rho u_r^2\\ (\rho E + p) u_r
\end{bmatrix}.
\label{eq:euler1d_spherical}
\end{equation}

Similarly, for problems with cylindrical symmetry (and assuming no circumferential flow), the solution satisfies
\begin{equation}
\begin{aligned}
    \Big(\rho(\bm{x},t),& ~u(\bm{x},t),~v(\bm{x},t),~w(\bm{x},t),~E(\bm{x},t)\Big) \\
    & = \Big(\rho(r,z,t),~u_r(r,z,t)\dfrac{x}{r},~u_r(r,z,t)\dfrac{y}{r},~ w(r,z,t),~E(r,z,t)\Big),
\end{aligned}
\label{eq:cylindrical_trans}
\end{equation}
where $z$ is the axial coordinate, $r=\sqrt{x^2+y^2}$ is the radial coordinate, and $u_r$ the radial velocity. The resulting 2D governing equations are 
\begin{align}
\dfrac{\partial}{\partial t}
\begin{bmatrix}
\rho \\  \rho u_r \\ \rho w  \\ \rho E
\end{bmatrix}
+
\dfrac{\partial}{\partial r}
\begin{bmatrix}
\rho u_r \\ \rho u_r^2 + p \\ \rho u_r w \\ (\rho E + p) u_r
\end{bmatrix}
+
\dfrac{\partial}{\partial z}
\begin{bmatrix}
\rho w \\ \rho u_r w \\ \rho w^2 + p \\ (\rho E + p) w  
\end{bmatrix}
= -\dfrac{1}{r}
\begin{bmatrix}
\rho u_r \\  \rho u_r^2 \\ \rho u_r w \\ (\rho E + p) u_r 
\end{bmatrix}.
\label{eq:Euler_cylindrical}
\end{align}

The test case in \texttt{Tests/BubbleInertCollapseCao21} illustrates these models by simulating a spherically symmetric bubble dynamics problem using both a 1D spherical domain and a 2D cylindrical domain.

\subsection{Interface tracking and treatment}
\label{sec:interface_track}

The internal boundaries of $\Omega$ that separate different subdomains are referred to as interfaces. M2C distinguishes between two types:
\begin{itemize}
\item Massless interfaces that move passively with the flow. 
\item Inertial interfaces, which possess their own dynamics and are governed by additional equations of motion.
\end{itemize}

Massless interfaces are typically used in multi-material flow simulations to represent boundaries between immiscible fluids, such as those enclosing bubbles or droplets.  Inertial interfaces are used in fluid-structure interaction simulations to model the ``wetted'' surfaces of solid bodies and thin-walled structures, that is, the portions of their surfaces in contact with fluid. For this reason, they are also referred to as fluid-fluid and fluid-structure interfaces, respectively.

\subsubsection{Massless (fluid-fluid) interfaces}
\label{sec:interface_ff}

Let $\Omega_{m}$ and $\Omega_{n}$ denote two neighboring subdomains separated by a massless interface. M2C enforces the following  conditions on the interface $\partial\Omega_{m} \cap \partial\Omega_{n}$:
\begin{equation}
\left.
\begin{array}{l}
\Big(\lim\limits_{\bm{x}'\rightarrow\bm{x},~\bm{x}'\in\Omega_{m}} \bm{V}(\bm{x}',t) - \lim\limits_{\bm{x}'\rightarrow\bm{x},~\bm{x}'\in \Omega_{n}} \bm{V}(\bm{x}',t) \Big)\cdot \bm{n}(\bm{x},t) = 0, \\ 
\lim\limits_{\bm{x}'\rightarrow \bm{x},~\bm{x}'\in \Omega_{m}} p(\bm{x}',t) = \lim\limits_{\bm{x}'\rightarrow \bm{x},~\bm{x}'\in \Omega_{n}} p(\bm{x}',t),
\end{array} 
\right.
\forall \bm{x}\in \partial\Omega_{m} \cap \partial\Omega_{n},~t\geq 0,
\label{eq:interface_cond}
\end{equation}
where $\bm{n}$ is the normal direction to the interface. In other words, the normal component of velocity and the pressure are continuous across the interface, while other state variables (e.g., density and internal energy) may exhibit finite jumps.

The motion of massless interfaces is tracked using the level set method, which allows for large deformations and topological changes such as merging and separation of subdomains. M2C implements a generalized version of the method that supports multiple types of massless interfaces that cannot merge (see examples in Sec.~\ref{sec:VandV_hypervelocity} and Sec.~\ref{sec:exmp_hypervelocity}). For a problem with $M$ subdomains bounded by massless interfaces, M2C solves $M-1$ level set equations of the form:
\begin{equation}
\frac{\partial \phi^{(m)} \left( \bm{x}, t \right)}{\partial t} + \bm{V} \cdot \nabla \phi^{(m)} = 0,\quad m=1,2,\cdots M-1,
\label{eq:levelset}
\end{equation}
where $\phi^{(m)}(\bm{x}, t)$ is a level set function representing the signed distance to the boundary of $\Omega_{m}$. At any time $t\geq 0$, the boundary is implicitly given by the zero level set:
\begin{equation}
\partial\Omega_{m}(t)=\{\bm{x}\in\Omega,~\phi^{(m)}(\bm{x},t)=0\}.
\end{equation}

All level set equations share the same velocity field $\bm{V}$, which prevents spurious overlap or separation of subdomains.

Initially (at $t = 0$), each level set function is constructed to satisfy the signed distance property:
\begin{equation}
|\nabla \phi| = 1.
\label{eq:signed_distance_property}
\end{equation}
However, this property is generally not preserved during the evolution governed by Eq.~\eqref{eq:levelset}, even if the equation is solved exactly. M2C can restore the property using a reinitialization procedure, performed at user-specified time intervals. In this procedure, M2C solves the following PDE to steady state \cite{sussman1994level}:
\begin{equation} 
\dfrac{\partial\phi}{\partial \tilde{t}} + S(\phi_0)(|\nabla\phi| - 1) = 0,
\label{eq:reinit}
\end{equation}
where $\tilde{t}$ is a fictitious time, $\phi_0$ is the pre-reinitialization level set function, and $S(\phi_0)$ is a smoothed sign function defined as
\begin{equation}
S(\phi_0) = \frac{\phi_0}{\sqrt{\phi_0^2 + \varepsilon^2}}.
\end{equation}
Here, $\varepsilon$ is a small regularization parameter, typically set to the minimum mesh element size. The steady-state solution of \eqref{eq:reinit} satisfies \eqref{eq:signed_distance_property} and serves as a new initial condition for further integration of the level set equation \eqref{eq:levelset}.

\subsubsection{Inertial (fluid-structure) interfaces}
\label{sec:interface_fs}

In this case, M2C enforces the following kinematic and dynamic conditions on the interface
\begin{equation}
\begin{aligned}
    & \bm{V} \cdot \bm{n} = \bm{V}_S \cdot \bm{n}, \\
    & p \cdot \bm{n} = \bm{\sigma}_S \cdot \bm{n},
\end{aligned}
\label{eq:bc_FSI}
\end{equation}
where $\bm{V}_S$ is the velocity of the interface, and $\bm{n}$ denotes its  normal direction. $\bm{\sigma}_S$ is the Cauchy stress tensor of the structure. The kinematic condition ensures that the normal component of the fluid velocity matches that of the structural velocity, preventing fluid penetration through the interface. The dynamic condition enforces the balance of forces at the interface.

Unlike fluid-fluid interfaces, the location and velocity of fluid-structure interfaces are not computed by M2C. Instead, they are supplied either through input files or by a coupled structural dynamics solver. M2C uses an embedded boundary method to track the interface location relative to the computational mesh; further details are provided in Sec.~\ref{sec:num_FSI}.

\subsection{Laser radiation}
\label{sec:laser_rad}

M2C can simulate laser absorption in a fluid flow by coupling the energy equation in \eqref{eq:NSEquation} with a laser radiation equation, as described in \cite{zhao2023simulating} and \cite{zhao2024vapour}:
\begin{equation}
\nabla\cdot(L\bm{s}) = \nabla L \cdot \bm{s} + (\nabla \cdot \bm{s})L = - \mu_\alpha L,
\label{eq:laser_eq}
\end{equation}
where $L = L(\bm{x})$ is laser irradiance (dimension: [mass][time]\textsuperscript{-3}) at position $\bm{x}\in \Omega_\text{L}$, and $\bm{s}= \bm{s}(\bm{x})$ denotes the laser beam direction at this position. $\Omega_\text{L}\subset\Omega$ represents the region in the computational domain that is exposed to laser. $\mu_{\alpha}$ denotes the laser absorption coefficient, which depends on the laser's wavelength, the fluid material, and  temperature. In M2C, $\mu_{\alpha}$ is defined to be a linear function of temperature $T$ for each material, i.e.,
\begin{equation}
\mu_{\alpha} = c(T - T_0) + \mu_{\alpha,0},
\end{equation}
where $c$, $T_0$, and $\mu_{\alpha,0}$ are user-specified parameters.

The laser propagation direction $\bm{s}$ does not need to be a constant. Instead, M2C allows the user to model parallel, diverging, and converging laser beams by defining $\bm{s}$ as:
\begin{equation}
    \bm{s} (\bm{x}) = 
    \begin{cases}
     \bm{s}_0 & \text{(parallel beam)},\\
     (\bm{x}_a - \bm{x}) / \left| \bm{x}_a - \bm{x} \right| & \text{(focusing beam)},\\
     (\bm{x} - \bm{x}_a) / \left| \bm{x} - \bm{x}_a \right| & \text{(diverging beam)},
    \end{cases}
    \label{eq:laser_direction}
\end{equation}
where $\bm{s}_0$ denotes the prescribed laser beam direction for a parallel beam.  $\bm{x}_a$ denotes the focal point for focusing and diverging beams.

The laser irradiance at the source boundary of $\Omega_\text{L}$ must be provided by the user as an input. The source boundary is assumed  to be a circular region with center $\bm{x}_0$ and radius $r_0$, representing either the actual laser source or a cross-section of the beam where irradiance is known or estimated. M2C supports both uniform and Gaussian irradiance distributions on the source boundary (the latter commonly referred to as a Gaussian beam~\citep{welch2011optical}):
\begin{equation}
L(\bm{x},t) = 
\begin{cases}
    \dfrac{P_0(t)}{\pi r_0^2},   & \text{(uniform)},  \\
    \dfrac{2P_0(t)}{\pi w_0^2} \exp\Big( \frac{-2 \big|\bm{x}-\bm{x}_0\big|^2}{w_0^2} \Big), & \text{(Gaussian)}, 
\end{cases}
\label{eq:Gaussian_beam_eq}
\end{equation}
where $w_0$ is the waist radius of Gaussian beam. $P_0(t)$ is the time-dependent laser power function defined by user through an input file (see example in \texttt{Tests/LaserCavitationZhao24}).

The radiative heat flux $\bm{q}_r$ in the energy equation is obtained by integrating the laser irradiance over all directions, given by
\begin{equation}
\bm{q}_r = \int_{4\pi} L(\bm{x}) \delta(\bm{\hat{s}}-\bm{s}) \bm{\hat{s}} ~d\hat{\bm{s}} = L \bm{s},
\label{eq:laser_radiativeflux}
\end{equation}
where $\delta(\bm{\hat{s}}-\bm{s})$ is the Dirac delta function ensuring that $L$ exists only along $\bm{s}$, a property assumed due to the high directionality of the laser beam.

\subsection{Phase transition}
\label{sec:phase_trans}

M2C implements the latent heat reservoir method, proposed in~\cite{zhao2023simulating}, to model phase change from liquid to vapor (i.e., vaporization). The fundamental idea of this method is to introduce a new field variable \( \Lambda(\bm{x}, t) \), which tracks the intermolecular potential energy in the liquid phase. As heat is added to the liquid, its temperature increases until it reaches the vaporization temperature \( T_{\text{vap}} \). Beyond this point, any additional heat no longer increases the temperature but is instead stored in \( \Lambda \). At any point within the liquid subdomain, if $\Lambda$ reaches the material's latent heat of vaporization, $l$, a phase transition occurs, converting the liquid to its  vapor phase. The stored energy in \( \Lambda \) is then released and added to the internal energy of the vapor. This phase transition is assumed to occur instantaneously via an isochoric process. In the current implementation, both \( T_{\text{vap}} \) and \( l \) are treated as constant material properties.

The state variables before and after phase transition are related by
\begin{equation}
    \rho^+ = \rho^-, \quad e^+ = e^- + \Lambda^-, \quad \Lambda^+ = 0, \quad \phi^{+} = -\dfrac{\Delta x}{2},
    \label{eq:phi_correction}
\end{equation}
where the superscripts $-$ and $+$ indicate the variable's value before and after phase transition. $\phi$ is the level set function tracking the boundary of sudomain occupied by the vapor phase. $\Delta x$ denotes the local element size in the computational mesh. The pressure and temperature after phase transition are obtained from the EOS of the vapor phase, i.e.
\begin{equation}
p^+ = p_\text{vapor}(\rho^+, e^+), \quad T^+ = T_\text{vapor}(\rho^+, e^+).
\label{eq:phase_transition}
\end{equation}

\subsection{Ionization}
\label{sec:ionization}

When a material is subjected to extreme mechanical or thermal loads, its energy can exceed the ionization energy of the constituent atoms, resulting in partial or full ionization. This process produces a plasma, characterized by a mixture of free electrons and ions. M2C predicts both the onset and extent of ionization using the models developed in~\cite{islam2023fluid,islam2023plasma}.

M2C solves the generalized Saha equation to model ionization. For each element, this equation expresses the number density ratio between the $(r+1)$-th ion ($r=0,1,2,...$) and the $r$-th ion as a function of temperature and plasma density (i.e., number density of electrons). Specifically,
\begin{equation}
\dfrac{n_{r+1,j}  n_e}{n_{r,j}} = \dfrac{2U_{r+1,j}}{U_{r,j}}\left[\dfrac{2\pi m_e k_B T}{h^2}\right]^{\frac{3}{2}} \text{exp}\left(-\dfrac{I_{r,j}^{\text{eff}}}{k_B T} \right),\quad r = 0, 1, ..., Z_j-1,~j=1,2,...,J,
\label{eq:Saha}
\end{equation}
where $r$ and $j$ index the ionization stage and chemical element, respectively. The atomic number of species $j$ is denoted by $Z_j$ (e.g., $Z_j = 8$ for oxygen), and $J$ is the total number of distinct elements present in the plasma mixture. The number density of the $r$-th ionization state of species $j$ is represented by $n_{r,j}$, while $n_e$ is the electron number density. The plasma is assumed to be in local thermodynamic equilibrium (LTE) under high-temperature, high-density conditions. The constants $m_e$, $k_B$, and $h$ are the electron mass ($9.11 \times 10^{-31}~\text{kg}$), Boltzmann constant ($1.38 \times 10^{-23}~\text{J/K}$), and Planck constant ($6.63 \times 10^{-34}~\text{J}\cdot\text{s}$), respectively. $U_{r,j}$ and $U_{r+1,j}$ denote the internal partition functions of the $r$-th and $(r+1)$-th ionization states of species $j$. The effective ionization energy, $I_{r,j}^{\text{eff}}$, accounts for modifications due to plasma interactions and will be defined in a subsequent section.

To have a well-posed problem, the Saha equation must be coupled with conservation principles, particularly that of charge and nuclei ~\cite{zaghloul_2004}. The conservation of charge is given by

\begin{equation}
   n_e = \sum_{j = 1 }^{J} \sum_{r = 1 }^{Z_j} r n_{r, j},
    \label{eq: con e1}
\end{equation}
and conservation of nuclei,
\begin{equation}
    n_{\text{H}} = \sum_{j = 1 }^{J} \sum_{r= 0 }^{Z_j} n_{r, j}.
    \label{eq: con h1}
\end{equation}
where $n_{\text{H}}$ is the number density of heavy particles (number of nuclei of atoms or ions). 

The non-ideality of the plasma is accounted for within this model through the depression of the ionization energy of each element in the plasma. As such  $I_{r,j}^{\text{eff}} = I_{r,j} - \Delta I_{r}$. Here, $I_{r,j}$ represents the ionization energy of the $r$-th charge state of the $j$-th element in an infinite vacuum, whereas $\Delta I_{r}$ represents the depression of this energy introduced by the ion's interactions with the other particles in the plasma. Therefore, for  an ideal plasma, $\Delta I_{r} = 0$, however, there are many models proposed in literature which attempt to model this phenomena. In general, $\Delta I_{r}$ can be given by,
\begin{equation}
    \Delta I_{r} = \frac{(r+1) e^2}{4 \pi \varepsilon_0 L_{r}},
    \label{eq:non_ideal_model}
\end{equation}
where $e$ is the charge of an electron, $\epsilon_0$ is the permittivity of free space, and $L_r$ is a model-dependent characteristic length. In particular, three models for $L_r$ have been implemented in M2C,
\begin{itemize}
    \item Griem model~\cite{griem1962high}: uses the Debye length to represent it, $L_r = \lambda_D$,
    \item Ebeling model~\cite{ebeling1976theory}: added a quantum correction via the de Broglie wavelength, $L_r = \lambda_D + \Lambda_B/8$,
    \item Griem-Fletcher Model~\cite{fletcher_thesis}: incorporates the Debye length and the average inter-nuclear spacing for metallic plasmas, $L_r = \sqrt{\lambda_D^2 + R_r^2}$.
\end{itemize}
Further details on these models and their applicability can be found in~\cite{zaghloul_2004,griem1962high,ebeling1976theory,fletcher_thesis}.

The partition function $U_{r,j}$ captures the effect of all accessible energy levels on the thermodynamic behavior of the $r$-th ionization state of species $j$ at temperature $T$. It is defined as
\begin{equation}
    U_r = \sum_{i = 1}^\infty g_{r,i}\exp\left(-\frac{E_{r,i}}{k_B T}\right),
    \label{eq:partition_function}
\end{equation}
where $g_{r,i}$ and $E_{r,i}$ denote the degeneracy and excitation energy of the $r$th ion at the $i$th energy level. In a plasma environment, ionization energy depression reduces the ionization threshold, effectively limiting the number of accessible excited states. Consequently, the summation in Eq.\eqref{eq:partition_function} is truncated at a maximum energy level $E_{n^*} \leq I_{r}^{\text{eff}} = I_{r} - \Delta I_{r}$. Thus, $U_r$ depends on both temperature ($T$) and ionization energy depression $(\Delta I_{r})$.

\section{Numerical methods}
\label{sec:numerical_method}

\subsection{Mesh generation}
\label{sec:mesh}

M2C provides built-in functionality for generating axis-aligned Cartesian grids. The user defines the computational domain, $\Omega$, by providing its upper and lower bounds along the $x$-, $y$-, and $z$-axes. The grid lines are generated along each axis independently. 
For a given axis (e.g., $x$), the user may choose between two options:
\begin{itemize}
  \item Uniform grid: Specify the number of elements, in which case M2C generates evenly spaced grid lines.
  \item Non-uniform grid: Provide a set of control points $\{(\tilde{x}_i, h_i)\}_{i=0}^n$, where each $\tilde{x}_i$ is a spatial location and $h_i$ is the preferred element size there. M2C then constructs a non-uniform grid that approximately matches the desired resolution by the method described in~\cite[Sec.~6.1]{quarteroni2009numerical}.
\end{itemize}

In the non-uniform case, a piecewise linear function $h(x)$ is constructed by linearly interpolating the control point data. The total number of elements, $N$, is computed by  
\begin{equation}
    \widetilde{N} = \int_{x_{\text{min}}}^{x_{\text{max}}} \dfrac{1}{h(x)}\,dx, \quad
    N = \max\left(1, \left\lfloor \widetilde{N} \right\rfloor \right),
    \label{eq:meshgen0}
\end{equation}
where $x_{\text{min}}$ and $x_{\text{max}}$ are the bounds of the domain, and $\lfloor \cdot \rfloor$ denotes the floor function.

The coordinates of the grid nodes $\{x_k\}_{k=0}^N$ are then determined by solving
\begin{equation}
    \frac{N}{\widetilde{N}} \int_{x_\text{min}}^{x_k} \dfrac{1}{h(x)}\,dx = k,
    \label{eq:meshgen}
\end{equation}
for each $k = 0, \dots, N$.  
Since $h(x)$ is piecewise linear, the integrals in \eqref{eq:meshgen0} and \eqref{eq:meshgen} can be evaluated analytically.

M2C uses the PETSc library to partition the mesh for distributed-memory parallelization. Each block of the mesh, assigned to one processor, has a rectangular shape. Ghost elements are used to exchange data between neighboring blocks, and to enforce boundary conditions at the fixed external boundary of the computational domain.

\subsection{Spatial discretization of governing equations}
\label{sec:num_fluid}

The governing equations~\eqref{eq:NSEquation} are discretized on the mesh using the finite volume method. M2C adopts a cell-centered formulation, in which control volumes (also referred to as cells) are defined as the elements of the primal grid described in Sec.~\ref{sec:mesh}. In general, field variables are stored at the geometric centers of the control volumes.

Integrating the governing equations within each control volume $C_i$ yields
\begin{equation}
\frac{\partial \bm{W}_i}{\partial t} = - \frac{1}{|C_i|} \sum_{j \in \text{Nei}(i)} \int_{\partial C_{ij}} \mathcal{F}(\bm{W}) \cdot \bm{n}_{ij} dS + \frac{1}{|C_i|} \int_{C_i} (\nabla \cdot \mathcal{G} (\bm{W}) )d\bm{x} + \mathcal{S}_i,
\label{eq:NS_discretized}
\end{equation} 
where $\bm{W}i$ denotes the average of $\bm{W}$ within $C_i$, and $| C_i |$ is the volume of the control volume. The set $\text{Nei}(i)$ contains the indices of control volumes adjacent to $C_i$, and $\partial C_{ij} = \partial C_i \cap \partial C_j$ denotes the common interface between $C_i$ and $C_j$. The unit normal vector $\bm{n}_{ij}$ points from $C_i$ to $C_j$, and $\mathcal{S}_i$ denotes average of the source term within $C_i$. Because of the Cartesian grid, $\bm{n}_{ij}$ is in the (positive or negative) $x$, $y$, or $z$ direction.

The surface integral of the advective flux, $\mathcal{F}(\bm{W})$, is approximated using the FIVER (FInite Volume method with Exact multi-material Riemann problem solvers) approach. In general,
\begin{equation}
\int_{\partial C_{ij}} \mathcal{F}(\bm{W}) \cdot \bm{n}_{ij} dS \approx  F_{ij},
\label{eq:advec_flux}
\end{equation}
where $F_{ij}$ denotes the FIVER approximation.

\begin{itemize}
\item If the two adjacent control volumes $C_i$ and $C_j$ are not separated by any material interface, $F_{ij}$ is computed using  the conventional monotonic upstream-centered scheme for conservation laws (MUSCL). In this case,
\begin{equation}
    F_{ij} = |\partial C_{ij}  | \Phi(\bm{W}_{ij}, \bm{W}_{ji}, \bm{n}_{ij}, \text{EOS}),
\end{equation}
where $|\partial C_{ij}  |$ denotes the area of $C_{ij}$, and $\Phi$ is a numerical flux function. M2C provides three options for $\Phi$: the local Lax–Friedrichs flux (also known as the Rusanov flux~\cite{rusanov1970difference}), Roe–Pike flux~\cite{hu2009hllc}, and HLLC flux~\cite{toro2013riemann} (see \ref{app:flux_func}). These fluxes are compatible with general EOS in the form described in Sec.~\ref{sec:EOS}. 

The variables $\bm{W}_{ij}$ and $\bm{W}_{ji}$ represent the reconstructed state vectors on either side of the interface $\partial C_{ij}$. M2C currently performs linear reconstruction within each control volume, with slopes limited by a slope limiter function. Available limiter options include the Monotonized Central-difference (MC) limiter, the Van Albada limiter, and a modified version of the Van Albada limiter (see~\ref{app:limiter}).

M2C allows reconstruction to be performed on the primitive state variables $(\rho,~\bm{V},~p)$, conservative state variables $(\rho,~\rho\bm{V},~\rho E)$, or the characteristic variables associated with either set. The definitions of characteristic variables and their reconstruction procedures in M2C follow the descriptions in~\cite{rodionov2019artificial}. The functions that convert between different variable formats are implemented in \texttt{FluxFcnBase.h}.  Due to the nonlinear nature of the governing equations, these four reconstruction options are not equivalent. There does not appear to be a clear consensus as to which approach performs best~\cite{van2006upwind,rodionov2019artificial,smith2007comparison,MIYOSHI2020109804}. Examples testing different numerical flux functions, slope limiters, and reconstruction variables can be found in \texttt{Tests/RiemannProblems}.

\item If $C_i$ and $C_j$ are separated by a massless (fluid–fluid) material interface, a one-dimensional (1D) bimaterial Riemann problem is constructed along the edge connecting their centers:
\begin{equation}
\frac{\partial \bm{w}}{\partial \tau} + \frac{\partial \bm{f}(\bm{w})}{\partial \xi} = 0, \quad \text{with} \quad \bm{w}(\xi,0) = 
\begin{cases}
&\bm{w}_i, \qquad \text{if} \ \xi \leq 0, \\
&\bm{w}_j, \qquad \text{if} \ \xi > 0.
\end{cases}
\label{eq:1DRiemannProblem}
\end{equation}
where $\tau$ is the local time variable, and $\xi$ is a spatial coordinate aligned with the interface normal $\bm{n}{ij}$. The initial states $\bm{w}_i$ and $\bm{w}_j$ are the projections of the reconstructed state variables, $\bm{W}_{ij}$ and $\bm{W}_{ji}$, on the $\xi$ axis. $\bm{f}$ denotes the advective flux function in 1D. Equation~\eqref{eq:1DRiemannProblem} is essentially the 1D Euler equations with piecewise constant initial conditions.  It admits a self-similar solution that satisfies the interface conditions in~\eqref{eq:interface_cond}. The solution states immediately adjacent to the interface are first mapped back to 3D, yielding $\bm{W}^{*}_{ij}$ and $\bm{W}^{*}_{ji}$, which are subsequently used in the same numerical flux function $\Phi$ to compute the interfacial fluxes:
\begin{align}
    F_{ij} &= | \partial C_{ij} | \Phi(\bm{W}_{ij}, \bm{W}^{*}_{ij}, \bm{n}_{ij}, \text{EOS}^{(i)}), \\
    F_{ji} &= | \partial C_{ij} | \Phi(\bm{W}_{ji}, \bm{W}^{*}_{ji}, \bm{n}_{ij}, \text{EOS}^{(j)}).
\end{align}

\item If $C_i$ and $C_j$ are separated by an inertial (fluid–structure) material interface, a one-sided Riemann problem is constructed on each side of the interface that lies within an active subdomain (i.e., not occluded by the solid structure):
\begin{equation}
\frac{\partial \bm{w}}{\partial \tau} + \frac{\partial \bm{f}(\bm{w})}{\partial \xi} = 0, \quad  \ \xi \leq 0, \quad \text{with} \quad \bm{w}(\xi,0) = \bm{w}_i, \quad v(0,\tau) = v_s.
\label{eq:1DHalfRiemannProblem}
\end{equation}

The boundary condition $v(0,\tau) = v_s$ prescribes the normal velocity of the interface at its intersection with the edge connecting the centers of $C_i$ and $C_j$. This interface velocity $v_s$ is either specified directly or computed by a coupled structural dynamics solver.  As in the fluid–fluid case, the solution of this one-sided Riemann problem provides the interfacial fluid state, which is then used in the numerical flux function to evaluate the flux across $\partial C_{ij}$.
\end{itemize}

In practice, solving Riemann problems with complex EOS can be computationally expensive due to the nested loops involved. To mitigate this issue, M2C implements the acceleration techniques proposed in~\cite{MA2023112474, ma2024data}. These methods do not introduce new approximations or require additional user input.
\vspace{2mm}

The diffusive fluxes in \eqref{eq:NS_discretized} ($\mathcal{G}(\bm{W})$, excluding the radiative flux term) are integrated also using a finite volume method as follows:
\begin{equation}
    \int_{C_i} \nabla \cdot \mathcal{G} (\bm{W}) d\bm{x} = \sum_{j \in \text{Nei}(i)} \int_{\partial C_{ij}} \mathcal{G}(\bm{W}) \cdot \bm{n}_{ij} dS = \sum_{j \in \text{Nei}(i)} | \partial C_{ij} | \mathcal{G}_{ij} \cdot \bm{n}_{ij},
\end{equation}
where $| \partial C_{ij} |$ is the area of $\partial C_{ij}$. $\mathcal{G}_{ij}$ is an approximation of $\mathcal{G}(\bm{W})$ at $\partial C_{ij}$, determined based on the specific terms involved. For the viscous term, any required quantities are reconstructed by linear interpolation of the values stored at the centers of $C_i$ and $C_j$. The heat diffusion term follows a similar interpolation method with energy balance enforced at the interface, as detailed in \cite{zhao2024vapour}. The treatment of radiative fluxes will be discussed separately in Sec.~\ref{sec:num_FLI}.

\subsection{Time integration}
\label{sec:time_integ}

After spatial discretization, Eq.~\eqref{eq:NS_discretized} becomes a system of ordinary differential equations in time. M2C provides  several time integration methods, including the second-order and third-order Runge-Kutta TVD schemes~\cite{gottlieb1998total}, which offer good accuracy while suppressing spurious oscillations. For rapid testing, the first-order forward Euler method is also available.

For example, the third-order Runge-Kutta method implemented in M2C proceeds as follows
\begin{equation}
\begin{aligned}
\bm{W}^{(1)} &= \bm{W}^{n} + \Delta t \bm{R}(\bm{W}^{n}),\\
\bm{W}^{(2)} &= \dfrac{3}{4} \bm{W}^{n} + \dfrac{1}{4} \bm{W}^{(1)} + \dfrac{1}{4} \Delta t \bm{R}(\bm{W}^{(1)}),\\
\bm{W}^{n+1} &= \dfrac{1}{3} \bm{W}^n + \dfrac{2}{3} \bm{W}^{(2)} + \dfrac{2}{3} \Delta t \bm{R}(\bm{W}^{(2)}),
\end{aligned}
\end{equation}
where $\bm{W}^{n}$ and $\bm{W}^{n+1}$ denote the numerical solution at consecutive time steps, and $\bm{W}^{(1)}$, $\bm{W}^{(2)}$ are intermediate stage variables. The operator $\bm{R}(\cdot)$ represents the numerical approximation of the right-hand side of \eqref{eq:NS_discretized}. The time step size $\Delta t$ may be specified as a constant or computed dynamically based on a user-specified Courant–Friedrichs–Lewy (CFL) number.

\subsection{Level set method}
\label{sec:num_levelset}

M2C implements two numerical methods for solving the level set equation~\eqref{eq:levelset}: a finite volume method based on the approach described in~\cite{main2014implicit}, and a finite difference method following~\cite{hartmann2008differential}, with a third-order upwind scheme. In both methods, the equation can be solved over the entire computational domain or restricted to a bandwidth around the zero level set. The latter is often referred to as the narrow-band level set method. The example cases in \texttt{Tests/LevelSetTests} illustrate and compare these two approaches.

The upwind scheme proposed in \cite{sussman1994level} is implemented in M2C to solve the level set reinitialization equation Eq.~\eqref{eq:reinit}. However, as noted in \cite{russo2000remark}, this method alone may not be able to restore $\phi$ to a signed distance function without altering the location of the interface (i.e., the zero level set). Therefore, special treatments are required for the first-layer nodes near the interface. M2C incorporates five approaches for reinitializing these first-layer nodes, including fixing $\phi$ values at first-layer nodes, applying corrections using neighbors across the interface (CR-1 and CR-2) as described in \cite{hartmann2008differential}, and iterative corrections with a forcing term introduced into Eq.~\eqref{eq:reinit} (HCR-1 and HCR-2) as proposed in \cite{hartmann2010constrained}.


\subsection{Fluid-structure coupling}
\label{sec:num_FSI}

M2C provides a two-way coupled framework for simulating fluid–structure interaction (FSI) problems~\cite{narkhede2025fluid}. It tracks fluid–structure interfaces using an embedded boundary method based on the collision-based algorithm from~\cite{wang2012computational}, modified to exploit the structured nature of Cartesian grids. In this method, each interface is assigned an \emph{intersector}, a utility that gathers geometric and topological information. The intersector identifies grid nodes occluded by the structure, determines edge-interface intersections, locates the closest points on the surface, and tracks regions swept by the interface over time.

At each time step, M2C updates the fluid-structure interface position and velocity based on data provided either by an external structural dynamics solver or directly by the user. The external solver can operate concurrently with M2C, allowing real-time data exchange. Currently, M2C employs a partitioned coupling approach with Aero-S \cite{aeros}. It can be coupled with other solvers in a similar manner. If the user chooses to directly prescribe the interface dynamics , they can program a \texttt{UserDefinedDynamics} object and link it to M2C using the \texttt{UserDefinedDynamicsCalculator} option in the input file. 

Once the interface position is updated, M2C invokes the intersectors to track the discrete interface within the computational grid. The numerical fluxes are then computed using the FIVER method, and the state variables of all fluid cells, including those affected by interface motion, are updated. The nodes swept by the embedded boundary are updated using a selective interpolation method that pulls data from neighboring cells. M2C computes the external load on the structure through the equilibrium interface condition (Eq.~\ref{eq:bc_FSI}). This load is computed and distributed over the fluid-structure intersection points using the interpolation and lofting method described in \cite{ho2020discrete}. 

The \texttt{Tests} directory currently includes two fluid-structure interaction simulations coupled with Aero-S. \texttt{Tests/ExplosionChamberNarkhede} presents an example featuring an explosion-induced shock wave interacting with a thin-walled containment chamber (also see Sec.~\ref{sec:exmp_FSI}). \texttt{Tests/UNDEXImplosion} provides another example in the context of underwater structural collapse induced by a near-field explosion. The directory also includes a test case in which the structural dynamics is prescribed by the user, located in \texttt{Tests/WaterEntry}.

\subsection{Laser-fluid coupling}
\label{sec:num_FLI}

In M2C, the laser radiation equation~\eqref{eq:laser_eq} is discretized using the same computational grid created for the Navier-Stokes equations. Integrating \eqref{eq:laser_eq} over an arbitrary cell $C_i$ yields
\begin{equation}
\sum\limits_{j\in Nei(i)} A_{ij}L_{ij} (\bm{s}_{ij}\cdot \bm{n}_{ij})  = - {|\, {C_i} \,|} \mu_{\alpha}(T_i)L_i,
\label{eq:FVM3D}
\end{equation}
where $L_{i}$ is the cell average of $L$ within $C_i$. $\bm{s}_{ij}$ represents the laser direction at $\partial C_{ij}$. $L_{ij}$ is the value of $L$ at $\partial C_{ij}$ and evaluated by the mean flux method, i.e.,
\begin{equation}
L_{ij} = 
\begin{cases}
\alpha L_i + (1-\alpha) L_j & \text{if } \bm{s}_{ij} \cdot n_{ij} \geq 0,\\
(1-\alpha) L_i + \alpha L_j & \text{if } \bm{s}_{ij} \cdot n_{ij} < 0,\\
\end{cases}
\label{eq:mean_flux_method}
\end{equation} 
where $\alpha\in [0.5, 1]$ is a numerical parameter. Substituting Eq.~\eqref{eq:mean_flux_method} into Eq.~\eqref{eq:FVM3D} yields a system of linear equations with the cell-averages of laser irradiance as unknowns. This system is solved iteratively using the Gauss-Seidel method.

In general, the computational grid does not contain a subset of nodes, edges, or elements that explicitly resolve the boundary of the laser radiation domain or the spatially varying laser propagation directions $\bm{s}(\bm{x})$. To address this issue, M2C implements an embedded boundary method proposed in \cite{zhao2023simulating}. This method involves the population of ghost nodes outside the side boundary of the laser domain using second-order mirroring and interpolation techniques. 

At each time step, after solving the laser radiation equation~\eqref{eq:laser_eq}, the divergence of the radiative flux, $\nabla \cdot \bm{q_r}$, is obtained by
\begin{equation}
 (\nabla \cdot \bm{q_r})_i  =  \nabla \cdot (L_i \bm{s}) = - \mu_{\alpha}(T_i) L_i,
\label{eq:NS_source_term}
\end{equation}
and then added to Eq.~\eqref{eq:NS_discretized} as
\begin{equation}
\int_{C_i} (\nabla\cdot \bm{q_r}) d\bm{x} = (\nabla\cdot \bm{q_r})_i {|\, {C_i} \,|}.
\label{eq:radiation_discretized}
\end{equation}

\subsection{Phase transition}
\label{sec:num_vaporize}

The method of latent heat reservoir described in Sec.~\ref{sec:phase_trans} is  implemented in M2C by updating the state variables at the end of each time step. The user specifies the phase transition model parameters in the input file, including the transition liquid material ID $m$, its corresponding vapor phase material ID $n$, the vaporization temperature $T_{\text{vap}}$, and the latent heat of vaporization $l$ for material $m$. For each control volume $C_i$ in $\Omega_m$, the liquid temperature $T_i$ is obtained from Eq.~\eqref{eq:Temperature_law} and compared with $T_{\text{vap}}$. 
\begin{enumerate}
    \item[(a)] If $T_i\leq T_{\text{vap}}$, and $\Lambda_i > 0$, the latent heat stored in the control volume is used to increase the local temperature up to $T_{vap}$. Specifically,
        \begin{equation}
            \begin{aligned}
             \Lambda_i &= \Lambda_i - \min (e_{vap}- e_i,~\Lambda_i), \\
             e_i &= e_i + \min (e_{vap}- e_i,~\Lambda_i), \\
             T_i &= T^{(m)}(\rho_i,~e_i),
            \end{aligned}
        \end{equation}
         where $e_{vap}$ denotes the internal energy corresponding to $T_{vap}$ and $\rho_i$, obtained using the EOS of the liquid phase (i.e., material $m$). $T_\text{liquid}(\cdot,\cdot)$ is the temperature law for the liquid phase. 
    \item[(b)] If $T_i > T_{\text{vap}}$, we reduce $T_i$ to $T_{\text{vap}}$, and move the excessive heat to the latent heat reservoir, $\Lambda_i$. The state variables is updated as
    \begin{equation}
     \begin{aligned}
        T_i &= T_{vap},\\
        \Lambda_i &= \Lambda_i + (e_i - e_{vap}), \\
        e_i &= e_i - (e_i - e_{vap}) = e_{vap},
     \end{aligned}        
    \end{equation}
\end{enumerate}
Following these operations, M2C compares $\Lambda_i$ with $l$ to determine if vaporization should occur in $C_i$. If $\Lambda_i \geq l$, this control volume undergoes phase transition, and the thermodynamic state variables are updated using~\eqref{eq:phi_correction} and \eqref{eq:phase_transition}.

The laser-fluid coupling and phase transition capabilities in M2C are demonstrated in \texttt{m2c/Tests\-/LaserCavitationZhao24} using a laser-induced thermal cavitation problem. This example is also presented in Sec.~\ref{sec:exmp_laser_cavitation}.

\subsection{Ionization equation solver}
\label{sec:num_FPI}

To compute the ionization state of a plasma, the Saha equation~\eqref{eq:Saha} is formulated as (cf.~\cite{zaghloul_2004}),
\begin{equation}
    Z_{\text{{av}}} = \sum_{j = 1}^{J} c_j  \left[\left(\sum_{r = 1}^{Z_j} \frac{r}{(Z_{\text{{av}}} n_{\text{H}})^r} \prod_{m = 1}^{r} f_{m, j}\right)  \middle/ \left( 1 + \sum_{r = 1}^{Z_j}\frac{\prod_{m = 1}^{r} f_{m, j}}{(Z_{\text{{av}}} n_{\text{H}})^r} \right) \right],
\label{eq:Zav}
\end{equation}
where $Z_{\text{{av}}}$ denotes the average charge number of the material (possibly a mixture of multiple elements), and $n_{\text{H}}$ the number density of heavy particles. $f_{m, j}$ is the Saha coefficient, given by (the right-hand side of Eq.~\eqref{eq:Saha})
\begin{equation}
    f_{m, j} = 2 \frac{U_{m, j}}{U_{m-1, j}}\left[\frac{2\pi m_e k_B T}{h^2}\right]^{\frac{3}{2}} \text{exp}\left(-\frac{I_{m-1,j}^{\text{eff}}}{k_B T} \right),\quad m=1,\cdots,Z_j.
    \label{eq: saha coeffecients}
\end{equation}

In the current implementation, the ionization model depends only on temperature and density, and is one-way coupled to the Navier-Stokes equations. Therefore, Eq.~\eqref{eq:Zav} is solved independently within each control volume of $\Omega_m$ at every time step, where $m$ denotes the material undergoing ionization. This equation is a 1D transcendental equation with $Z_{\text{av}}$ as the sole unknown and is solved using a safeguarded iterative root-finding method --- specifically, the TOMS748 algorithm~\cite{alefeld1995algorithm} from the Boost library.

To accelerate the solution process, $U_{r,j}$ is tabulated as functions of $exp(-1/T)$ at the beginning of the analysis. During each time step, its value is obtained using interpolation, with cubic B-splines as the default method. For non-ideal plasmas, the ionization energy shift $\Delta I_{r,j}$ depends on plasma properties, requiring an iterative solution for both $Z_{\text{av}}$ and the Debye length $\lambda_D$, following the method described in \cite{islam2025ionization}. Once Eq.\eqref{eq:Zav} is solved, the molar fractions of neutral atoms and each ionized state are determined by evaluating the contribution of free electrons to the average charge per heavy particle, as described in \cite{islam2023fluid}.

The ionization modeling capability in M2C is demonstrated in \texttt{m2c/Tests\-/HVImpactShafquatIslam\-/2D\_axisym\_HVI} through a hypervelocity impact test case, which is also discussed in Sec.~\ref{sec:exmp_hypervelocity}. Additionally, a standalone solver of the Saha equation is available, as described in Sec.~\ref{sec:VandV_ionization}.

\section{Solver verification and validation}
\label{sec:VandV}

\subsection{One-dimensional Riemann problems}
\label{sec:VandV_riemann}

The directory \texttt{Tests/RiemannProblems} contains 12 test cases where M2C is applied to solve one-dimensional Riemann problems, governed by the 1D Euler equations. These tests feature the propagation of shock waves and rarefactions due to discontinuities in initial state and thermodynamic relations across a massless material interface. In each test, the numerical solutions of pressure, density, and velocity are compared with reference solutions obtained from a standalone bimaterial Riemann problem solver, which is available in the GitHub repository~\cite{standaloneRiemann}. Test cases $1$-$7$, taken from~\cite{Rallu2009}, show that M2C achieves comparable accuracy to the original study for the same grid resolution. Cases $8$-$13$ test primarily the Mie-Gr\"{u}neisen (MG) equation of state and problems related to high-speed impact. 

For instance, Test 12 models the interaction between a solid material (soda lime glass) and air. At time $t=0$, the left half of the computational domain ($x<0$~\text{mm}) is occupied by air, while the right half ($x>0$~\text{mm}) is occupied by soda lime glass. The initial conditions are  
\begin{equation}
(\rho, u, p, \mathrm{EOS})=\left\{\begin{array}{ll}
\left(1.2\times10^{-6}~\text{g/mm$^3$}, 0.0~\text{mm/s}, 1.0\times10^{5}~\text{Pa}, \text{stiffened~gas}\right), & x\leq 0~\text{mm} \\
\left(2.204\times10^{-3}~\text{g/mm$^3$},  1.5\times10^6~\text{mm/s}, 1.0\times10^{5}~\text{Pa},  \text{Mie-Gr\"{u}neisen}\right), & x>0~\text{mm}.
\end{array}\right.
\label{eq:RiemannExample}
\end{equation}

Air is modeled using the Noble-Abel stiffened gas EOS. With parameters $\gamma=1.4$, $q=0$, $b=0$, and $p_c=0$, it degenerates to the a perfect gas. Soda lime glass is modeled using the Mie-Gr\"{u}neisen EOS, with parameter values $\rho_0 = 2.204\times10^{-3}~\text{g/mm$^3$}$, $c_0 = 2.22\times10^{6}~\text{mm/s}$, $s = 1.61$, and $\Gamma_0 = 0.65$~\cite{MA2023112474,ma2024data}. At any time $t>0$, density jumps by $7$ orders of magnitude across the material interface, from $0.00052~\text{kg}/\text{m}^3$ to $2203.98~\text{kg}/\text{m}^3$. 

\begin{figure}[H]
    \centering
    \begin{subfigure}[b]{80mm}
       \includegraphics[width=80mm,trim={0cm 0cm 0cm 0cm},clip]{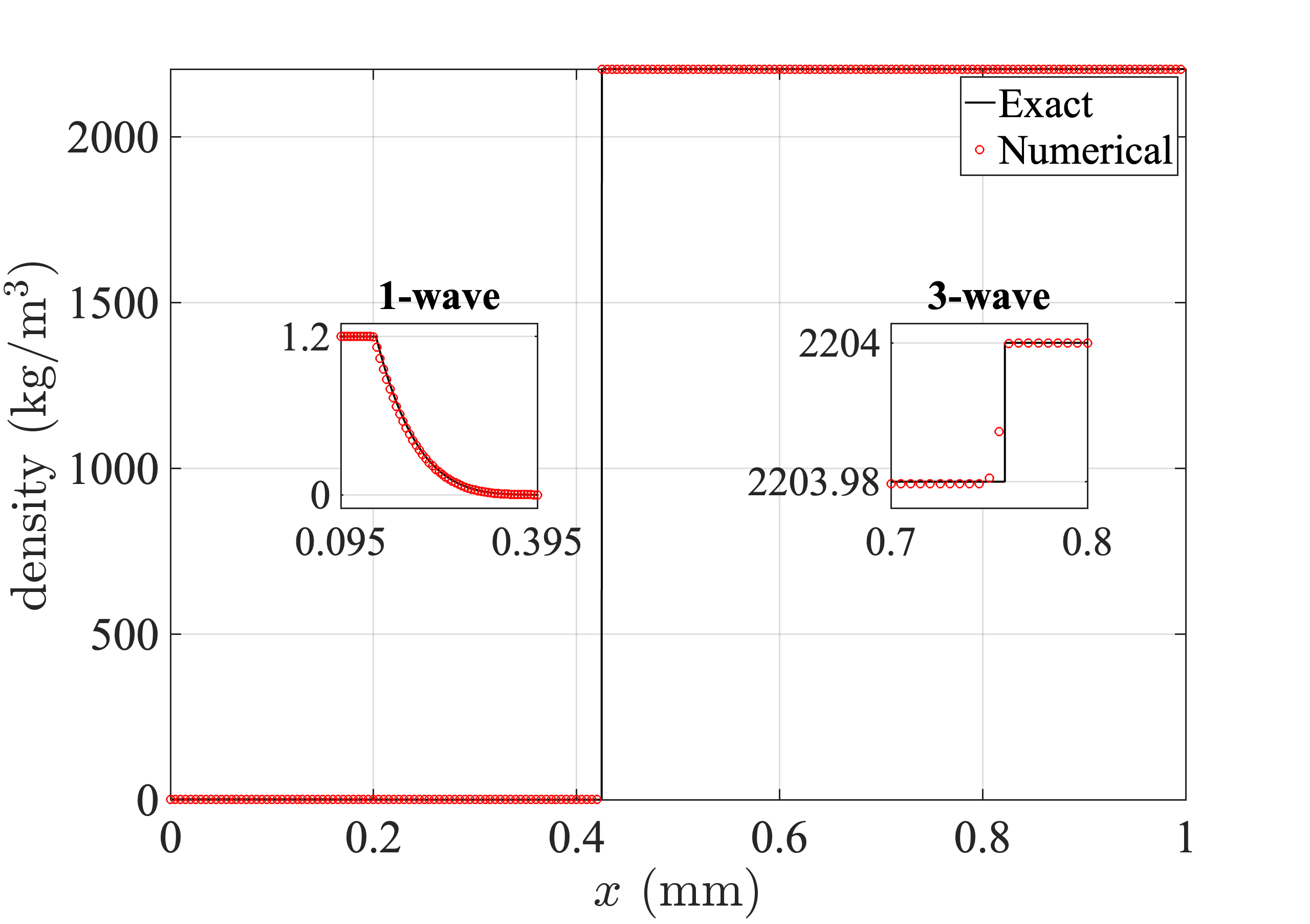}
          \caption{}
    \end{subfigure}
    \begin{subfigure}[b]{80mm}
           \includegraphics[width=80mm,trim={0cm 0cm 0cm 0cm},clip]{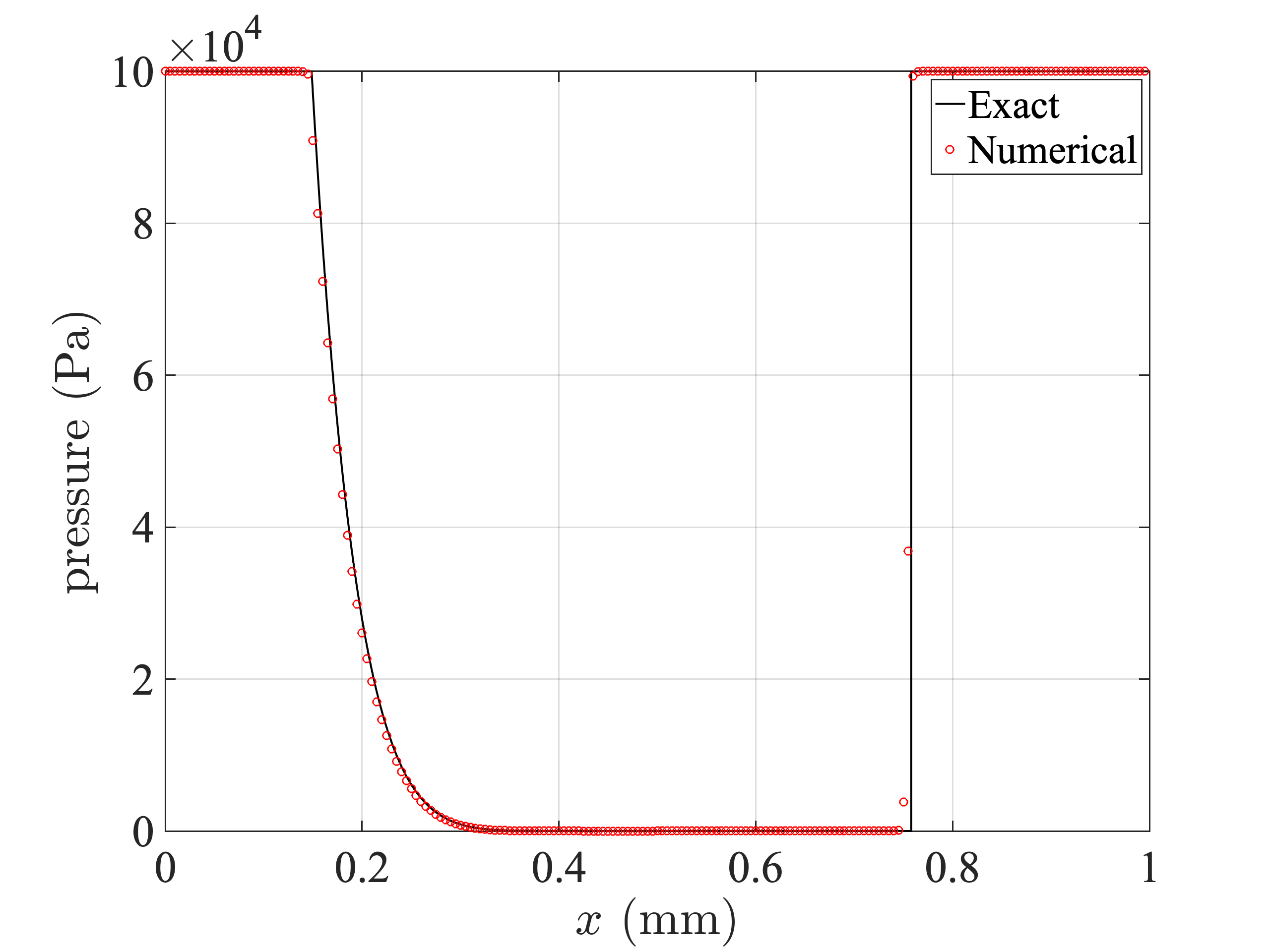}
          \caption{}
    \end{subfigure}
    \begin{subfigure}[b]{80mm}
       \includegraphics[width=80mm,trim={0cm 0cm 0cm 0cm},clip]{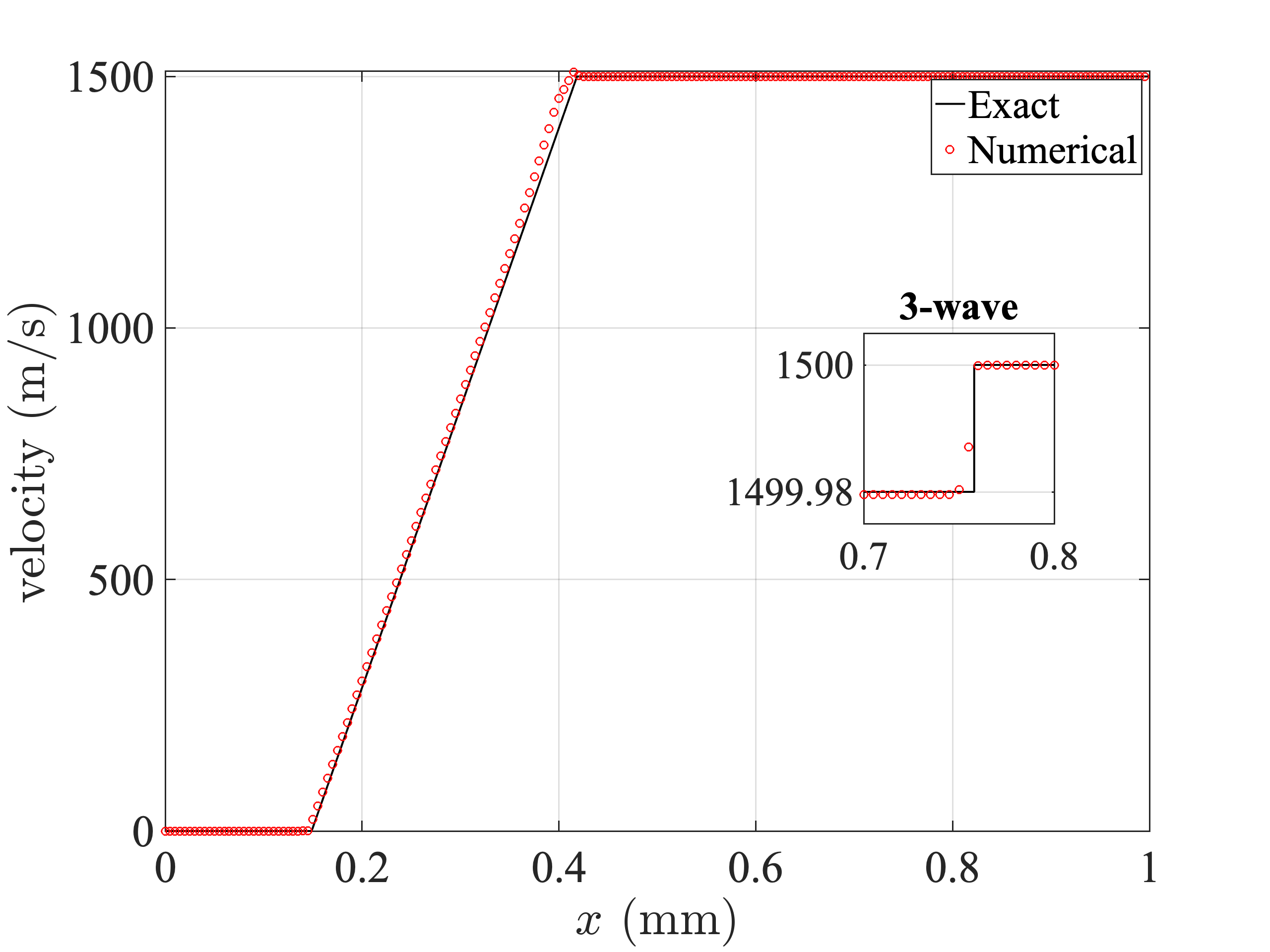}
      \caption{}
     \end{subfigure}
    \caption{A 1D bimaterial Riemann problem:~Density, pressure, and velocity distributions at $t=0.15~\mu$s.}
    \label{fig:1dCFD}
\end{figure}

The density, pressure, and velocity at $t = 0.15~\mu\text{s}$ are shown in Fig.~\ref{fig:1dCFD}, in comparison with the exact solution.  Subfigure (a) clearly shows the three-wave structure of the solution. Across the contact discontinuity (the 2-wave), the significant density jump is captured without numerical oscillations. Both the 1- and 3-waves are rarefactions, and the transitions in density, pressure, and velocity are captured accurately.

\subsection{Interface tracking}
\label{sec:VandV_levelset}

The directory \texttt{Tests/LevelSetTests} contains three benchmark problems for verifying the level set method implemented in M2C:
\begin{itemize}
\item Case 1: Rotation of a slotted disk \cite{hartmann2008differential},
\item Case 2: Vortex-induced deformation of a circle \cite{hartmann2010constrained}, and
\item Case 3: Merging and separation of two circles \cite{hartmann2008differential}.
\end{itemize}

In all three cases, the velocity field is prescribed at all times, so the level set equation is solved independently of the flow equations. To run these tests, M2C must be compiled with the preprocessor variable \texttt{LEVELSET\_TEST} set to \texttt{1}, \texttt{2}, or \texttt{3}, corresponding to the desired case. This can be done by adding  \texttt{add\_definitions(-DLEVELSET\_TEST=[case number])} to the \texttt{CMakeLists.txt} file. In this test mode, the Navier-Stokes equations solver is deactivated.

\begin{figure}[H]
\centering
\includegraphics[width = 0.99 \textwidth]{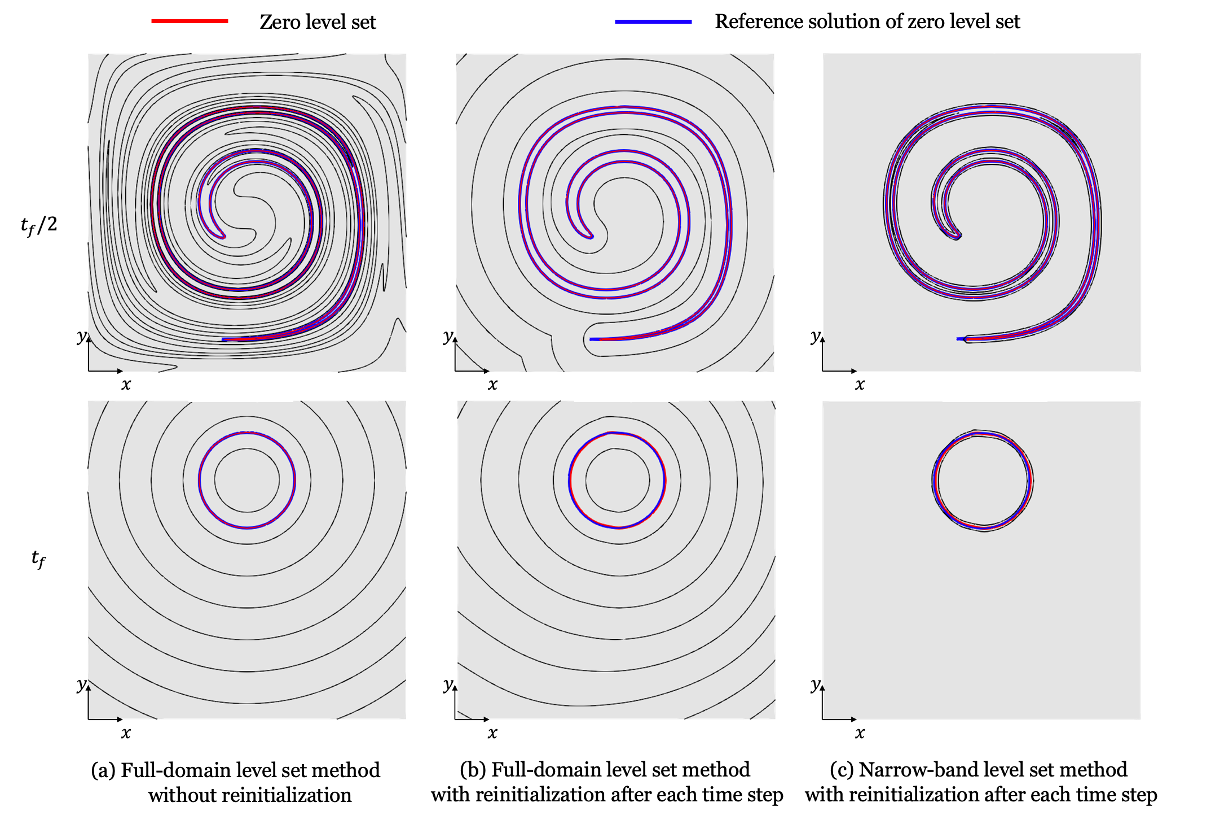}
\caption{Vortex-induced deformation of a circle. Red: zero level set from each method. Blue: reference solution computed using Method (a) on a $1024^2$ mesh. At $t_f = 8.0$, the interface returns to its original circular shape.}
\label{fig:votex_deform}
\end{figure}

For example, Case 2 applies a prescribed, divergence-free velocity field to deform an initially circular interface into a spiral, which then returns to its original shape at the final time $t_f = 8.0$. The simulation setup follows the configuration described in \cite{zhao2023simulating}. Figure~\ref{fig:votex_deform} shows simulation results on a $1024^2$-cell mesh using three approaches: (a) full-domain level set without reinitialization, (b) full-domain level set with reinitialization at every time step, and (c) narrow-band level set with reinitialization at every time step. All methods solve Eq.~\eqref{eq:levelset} using the finite difference method, with reinitialization performed by constraining first-layer nodes.

At $t = t_f/2$, all methods capture the over deformation of the interface. The zero level set obtained with Method (a) aligns with the reference solution, but the non-zero contours deviate, indicating a loss of the signed distance property in $\phi$. Methods (b) and (c), which incorporate reinitialization, preserve the signed distance property and produce consistent results. At $t = t_f$, all three methods are able to recover the original shape of the interface.


\subsection{Blast wave from gaseous detonation}
\label{sec:VandV_expansion}

This example simulates the blast wave generated by an open-air detonation. Initially, the gaseous explosion products (i.e., the burnt gas) are located within a spherical region of radius $r$, determined from the explosive charge’s mass and density. The initial state within this sphere are prescribed using the self-similar solution for spherical blast wave propagation (see \cite{narkhede2025fluid} for details). These conditions are provided to the M2C solver via a text file, which is specified in the main input configuration.

The surrounding air is initialized at ambient conditions, with a density of $1.177\times10^{-3}\text{kg/m}^3$ and a pressure of $100\text{Pa}$. The air is modeled as a perfect gas, while the burnt gas follows the JWL EOS to capture the high-explosive behavior. The interface between the burnt gas and air is tracked using the level-set method described in Sec.~\ref{sec:num_levelset}. The model setup and input files are located in the directory \texttt{Tests/DetonationExpansion}.

\begin{figure}
    \centering
    \includegraphics[width=\linewidth]{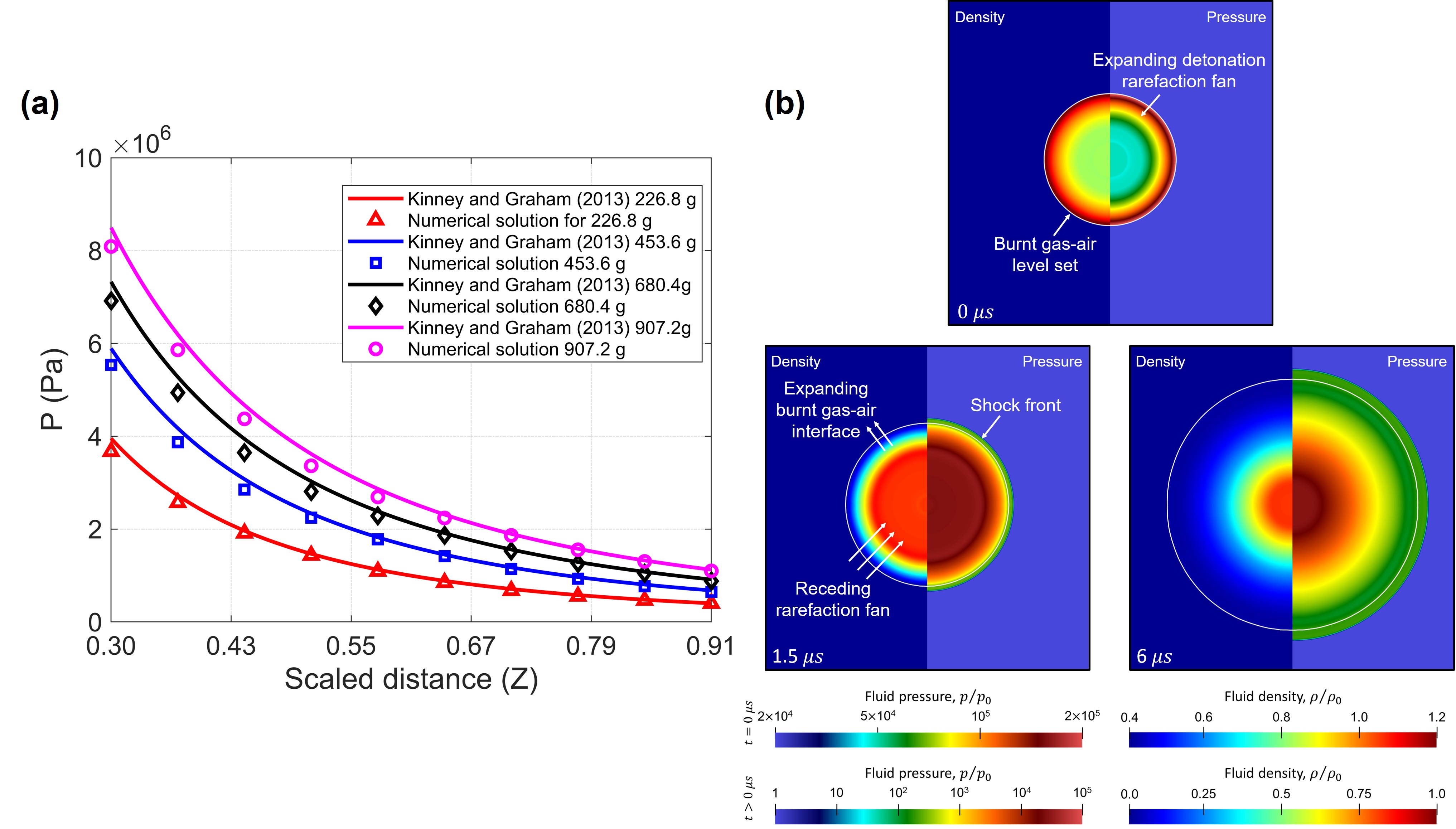}
    \caption{Blast wave from gaseous detonation: (a) Comparison of peak pressure values from simulation and semi-empirical prediction at probe locations. (b) Pressure and density snapshots from the $m = 0.2268~\text{kg}$ test case showing shock propagation and interface evolution.}
    \label{fig:burnt_gas_propagation}
\end{figure}

Four simulations with explosive charge masses of $m = 0.2268~\text{kg}$, $0.4536~\text{kg}$, $0.6804~\text{kg}$, and $0.9072~\text{kg}$ were carried out, with the TNT density fixed at $1630~\text{kg/m}^{3}$. In each test, ten pressure probes were uniformly distributed within the computational domain. Figure~\ref{fig:burnt_gas_propagation}(a) compares the maximum pressure values recorded at these probe locations with the predictions obtained from the semi-empirical equation provided by \cite{kinney_explosive_2013} for chemical explosion-induced overpressure. In all four test cases, the results are in reasonable agreement with the reference. Three density and pressure snapshots obtained with $m=0.2268~\text{kg}$ are shown in Fig.~\ref{fig:burnt_gas_propagation}(b). A shock wave is generated at the burnt gas - air interface due to the initial discontinuities, propagating outward into the ambient air. At the same time, a backward moving rarefaction fan interacts and diminishes the strength of the rarefaction wave produced during the detonation process. Driven by its high kinetic energy, the burnt gas pushes into the initially quiescent air, causing the material interface to expand outward. 

\subsection{Inertial collapse of spherical bubble in free-field}
\label{sec:VandV_collapse}

This example considers the collapse of a spherical bubble in an infinite liquid medium. The setup is based on the experiment reported in \cite{kroninger2010particle}. The simulation begins at the point of maximum bubble radius ($0.7469~\text{mm}$). The bubble content is modeled as a perfect gas ($\gamma=1.4$), with initial conditions: density $0.957\times10^{-3}\text{kg/m}^{3}$, velocity $0~\text{m}/\text{s}$, and pressure $100~\text{Pa}$.  The surrounding liquid is water, modeled using the stiffened gas EOS, with initial conditions:   density: $1000~\text{kg/m}^{3}$, velocity $0~\text{m}/\text{s}$, and pressure $1.01 \times 10^5 \text{Pa}$. Additional details can be found in \cite{cao2021shock}. The M2C input file and some sample results for this problem are also available in the directory \texttt{Tests/BubbleInertCollapseCao21}.

\begin{figure}[H]
    \centering
    \includegraphics[width=1\linewidth]{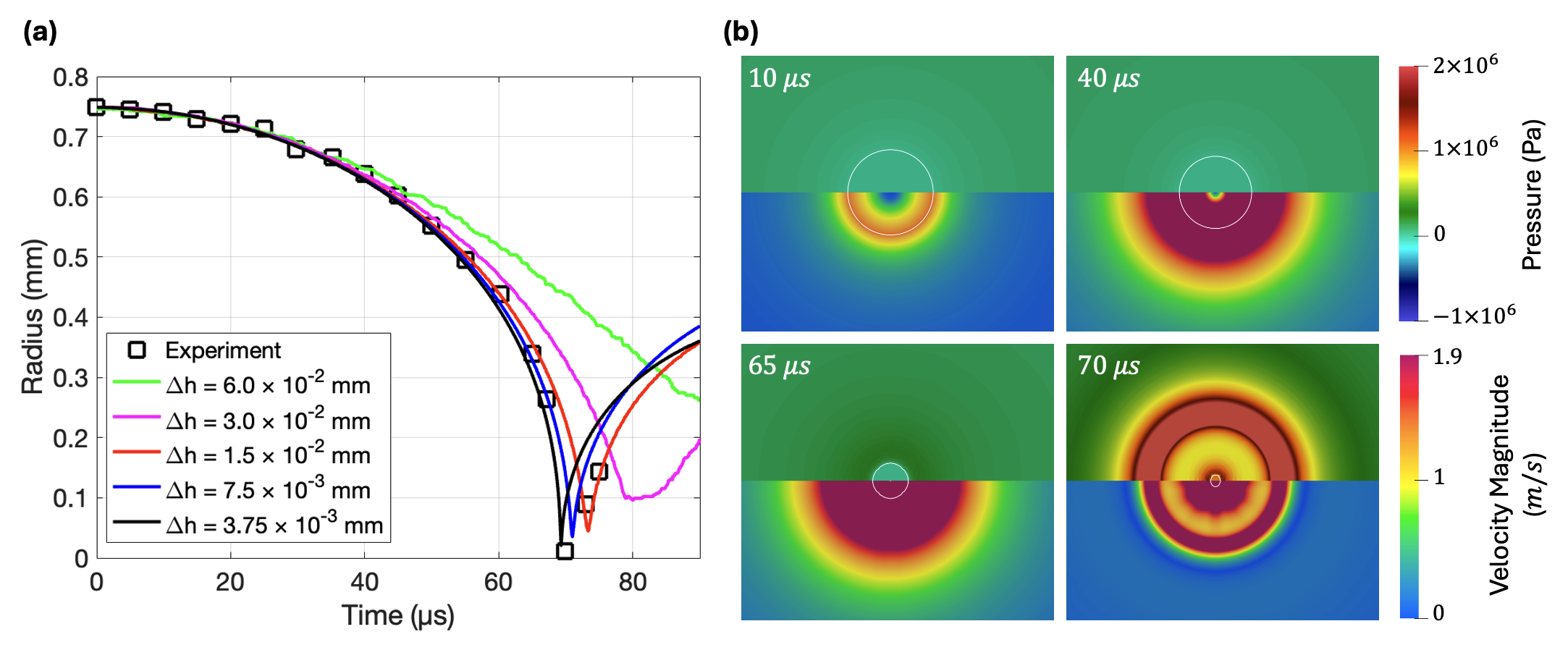}
    \caption{Collapse of a spherical bubble: (a) Time history of bubble radius. Five numerical solutions with different mesh resolutions ($\Delta h$) are shown, in comparison with experimental data from \cite{kroninger2010particle}. (b) Converged solution at four representative time instances during the collapse. Top: Pressure field. Bottom: Velocity magnitude. The white contour line denotes the bubble interface.}
    \label{fig:inertial_collapse}
\end{figure}

Figure~\ref{fig:inertial_collapse}(a) shows the time evolution of the bubble radius computed using five different mesh resolutions, with minimum element size ranging from $60~\mu\text{m}$ to $3.75~\mu\text{m}$. The figure shows that as the mesh is refined, the simulation results converge toward the experimental data. As the bubble rebounds, some discrepancy between the numerical and experimental results is observed. This may be due to simplified bubble content modeling.  Figure~\ref{fig:inertial_collapse}(b) shows snapshots of the pressure and velocity fields at four representative time instances. The bubble remains nearly spherical throughout the collapse process, and the simulation clearly captures the emission of a shock wave at the point of minimum volume.

\subsection{Shock-induced bubble collapse near a rigid wall}
\label{sec:VandV_collapse_wall}

\begin{figure}[H]
    \centering
    \includegraphics[width=1\linewidth]{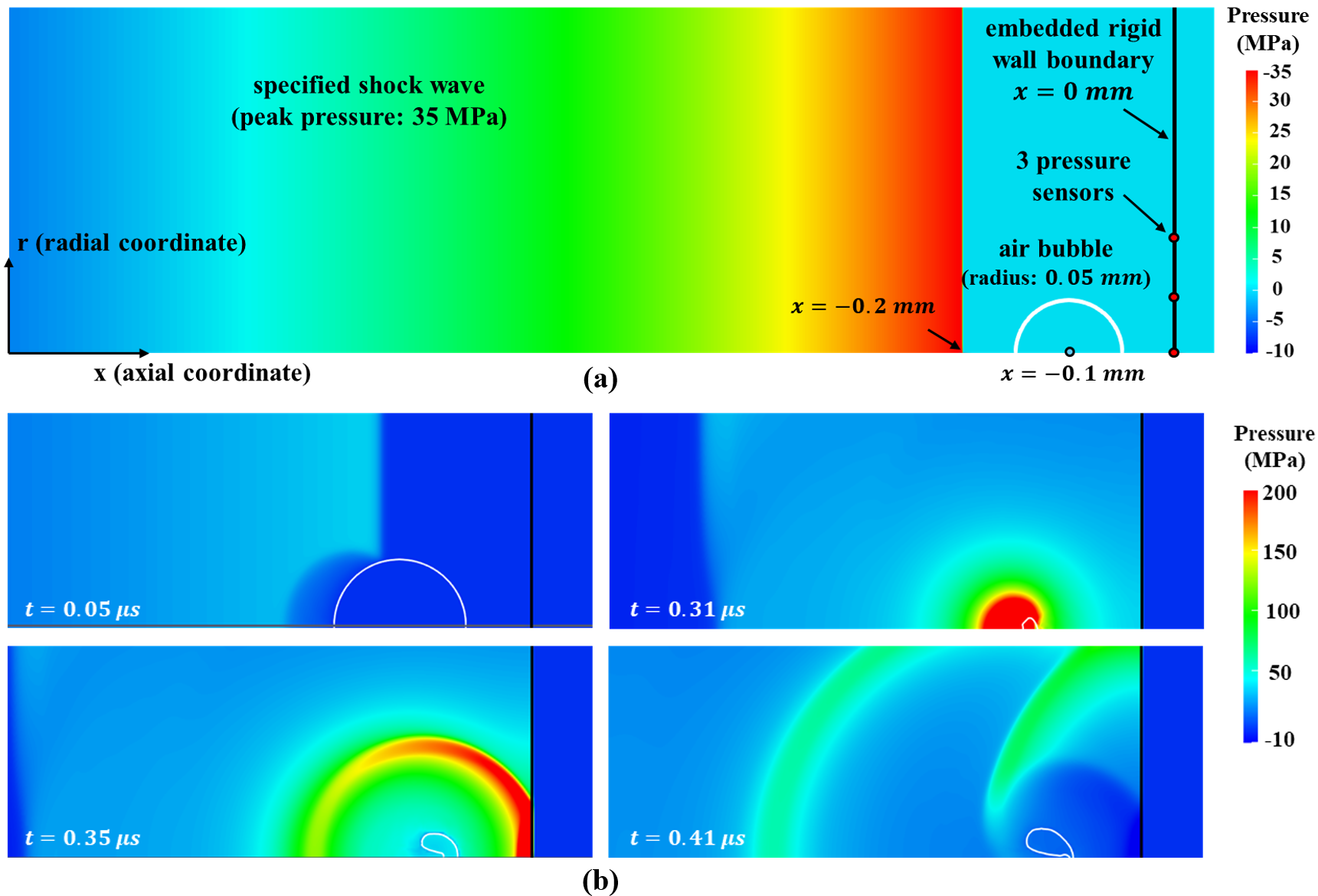}
    \caption{Shock-induced bubble collapse near a rigid wall: (a) Schematic of the problem setup and initial pressure distribution; (b) Pressure fields at four successive time instants."}
    \label{fig:shock_bubble_setup}
\end{figure}

This example problem was originally introduced by Johnsen and Colonius, analyzed using their research code~\cite{johnsen2008shock}. Later, comparable results were obtained using the Aero-F solver~\cite{wang2017multiphase,cao2021shock}. Figure~\ref{fig:shock_bubble_setup}(a) illustrates the problem setup. Initially, a spherical air bubble of radius $0.05$ mm is positioned $0.05$ mm away from a planar rigid wall (measured from the bubble's nearest point to the wall). The bubble contains air, with initial conditions: density $1.225 \times 10^{-6}\text{g}/\text{mm}^3$, velocity $0~\text{mm}/\text{s}$, and pressure $1.01 \times 10^5~\text{Pa}$.  The surrounding fluid medium is liquid water, with ambient conditions: density $1.0 \times 10^{-3}~\text{g}/\text{mm}^3$, velocity $0~\text{mm}/\text{s}$, and pressure $1.01 \times 10^5~\text{Pa}$. Both fluids are assumed to be compressible and inviscid. Air is modeled as a perfect gas with specific heat ratio $\gamma = 1.4$. Liquid water is modeled using the stiffened gas EOS, a simplified form of Eq.~\eqref{eq:NASG_EOS} with $e_c = 0$ and $b=0$.
The modeled parameters are $\gamma = 6.59$ and $p_c = 4.1\times 10^8~\text{Pa}$.

A planar shock wave propagating toward the bubble and the wall is prescribed by the time-dependent pressure function
\begin{equation}
p(t) = p_0 + 2p_s e^{-\alpha t} \cos\big(\omega t + \dfrac{\pi}{3}\big),
\end{equation}
with parameters $p_0 = 1.01 \times 10^5~\text{Pa}$, $p_s = 3.5 \times 10^7~\text{Pa}$, $\alpha = 1.48 \times 10^6~\text{s}^{-1}$, and $\omega = 1.21 \times 10^6~\text{s}^{-1}$. This temporal function is mapped onto the computational domain using the relation $t = - (x + x_s)/c_0$, where $x$ denotes the wave propagation axis, and $c_0$ is the speed of sound determined from the ambient water condition using the stiffened gas EOS. At time $t = 0$, the shock front is positioned at $x_s = -0.2$ mm, as indicated in Figure~\ref{fig:shock_bubble_setup}(a). The $x$-component of flow velocity within the shock profile is set by $u = (p - p_0)/(\rho_0 c_0)$, where $\rho_0 = 1.0 \times 10^{-3}~\text{g}/\text{mm}^3$ is the ambient water density. As time progresses, the shock wave propagates toward the bubble, ultimately inducing its collapse.

In the M2C simulation, the shock profile is prescribed via a user-defined initial condition file, which is compiled into a shared object and passed to the solver as a dynamically linked library. All necessary input files, a sample simulation log, and representative results are included in the source distribution under the directory \texttt{Tests/ShockBubbleCollapseWang17}. The simulation is performed on a 2D mesh with cylindrical symmetry enforced through the inclusion of geometric sink terms in the fluid governing equations. The minimum element size is $1.5\times 10^{-3}~\text{mm}$, consistent with the resolution used in \cite{wang2017multiphase}.  The planar wall is represented as a fixed embedded boundary, while the gas-liquid interface is captured by solving the level set equation. Additional numerical options and parameters are specified in the simulation input file: \texttt{Tests\-/ShockBubbleCollapseWang17\-/input.st}.

\begin{figure}[H]
    \centering
    \includegraphics[width=1\linewidth]{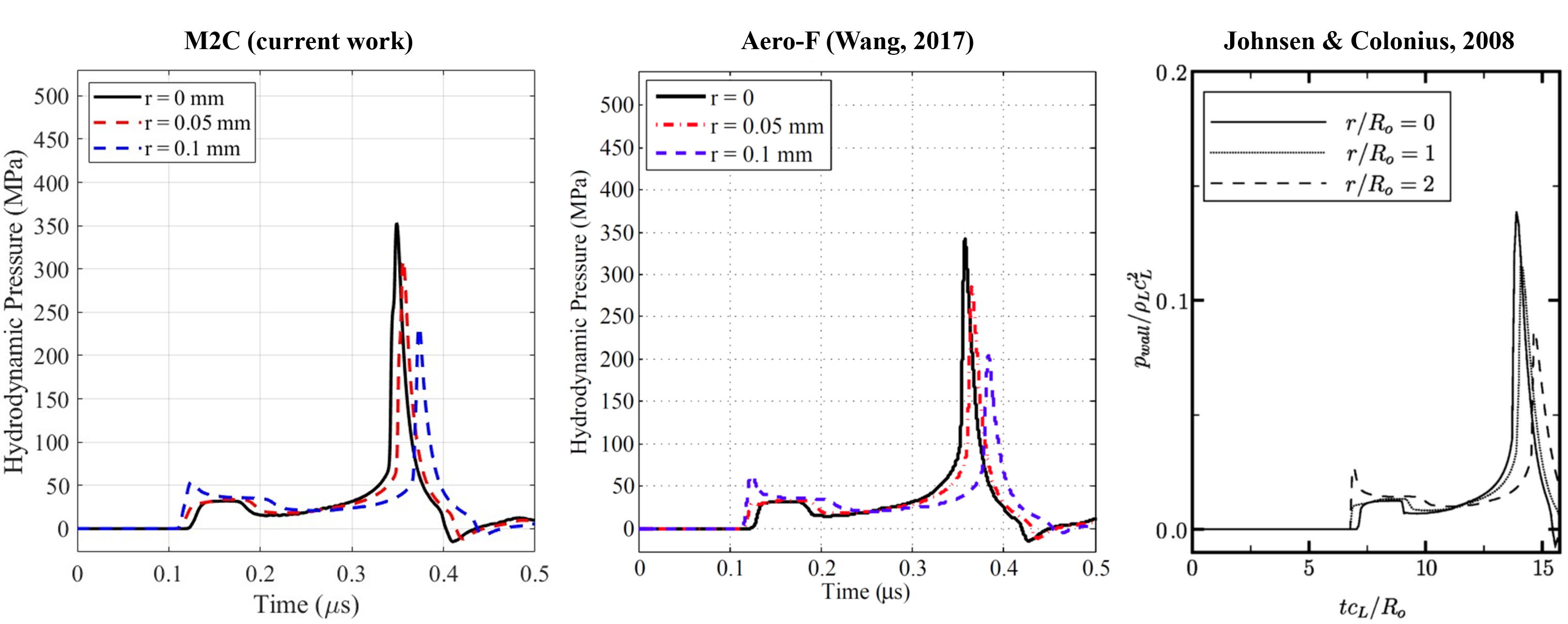}
    \caption{Shock-induced bubble collapse near a rigid wall: Comparison of pressure time-histories predicted by different solvers. The pressures from Johnsen and Colonius~\cite{johnsen2008shock} were nondimensionalized by $\rho_L c_L^2$. A dimensionalized value of $0.1$ is approximately  $2.7\times 10^9~\text{Pa}$ by our calculation.}
    \label{fig:shock_bubble_compare}
\end{figure}

Figure~\ref{fig:shock_bubble_setup}(b) presents the pressure field at four representative time instants, illustrating shock impact on the bubble, asymmetric bubble collapse, generation of a secondary shock, and bubble rebound. These results show strong qualitative agreement with those obtained using Aero-F (Figure 11 of \cite{wang2017multiphase}).  Furthermore, Figure~\ref{fig:shock_bubble_compare} compares pressure time-histories recorded at three sensor locations along the wall, as indicated in Figure~\ref{fig:shock_bubble_setup}(a). The M2C results closely match those presented in \cite{wang2017multiphase} and~\cite{johnsen2008shock}, with good agreement in both waveform and peak values.

\subsection{Laser absorption by multi-material fluid flow}
\label{sec:VandV_laser}

\begin{figure}[H]
\centering
\includegraphics[width = 0.99 \textwidth]{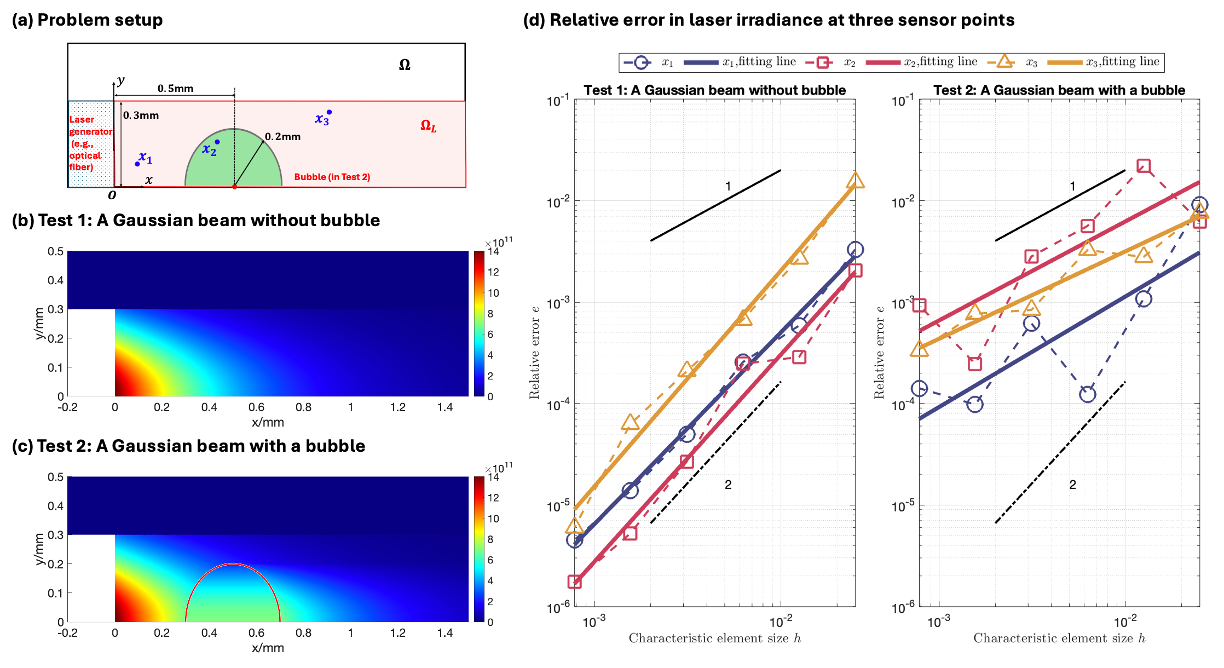}
\caption{Laser beam propagation in fluid media.
(a) Schematic of the problem setup. (b-c) Numerical solution of laser irradiance for Test 1 (without bubble) and Test 2 (with bubble) on the finest mesh. (d) Relative error in irradiance at three sensor points.}
\label{fig:laser_verification}
\end{figure}

The numerical methods for solving the laser radiation equation (Sec.~\ref{sec:num_FLI}) were analyzed analytically and numerically in recent work~\cite{zhao2023simulating}. It was shown that the mean flux method achieves second-order accuracy when $\alpha = 0.5$ in Eq.\eqref{eq:mean_flux_method}, and that the embedded boundary method preserves second-order accuracy. Two numerical tests are presented here, simulating  a Gaussian laser beam propagating in  single-material and multi-material fluid domains. Figure \ref{fig:laser_verification}(a) illustrates the problem setup.  Since the laser beam is collimated (i.e., neither focusing nor diverging), an exact solution to Eq.~\eqref{eq:laser_eq} can be derived and used as a reference. Simulations are conducted on a series of meshes with characteristic edge lengths ranging from $2.5\times 10^{-2}$ mm to $7.8\times 10^{-4}$ mm. In each simulation, the irradiance is recorded at three sensor locations, and the relative error at each sensor is computed with respect to the  exact solution.

Figure~\ref{fig:laser_verification}(d) shows the relative error versus mesh size. In Test 1, second-order convergence is observed at all sensor points, including the one adjacent to the embedded boundary ($x_3$), confirming that the embedded boundary method maintains second-order accuracy. In contrast, Test 2 shows only first-order convergence due to the discontinuity in the absorption coefficient $\mu_\alpha$ across the bubble interface. This discontinuity leads to a corresponding discontinuity in the irradiance field, thereby limiting the achievable convergence rate.

\subsection{Verification of ionization model implementation}
\label{sec:VandV_ionization}

The ionization module in M2C computes the solution to Eq.~\eqref{eq:Zav} based on local mass density ($\rho$), pressure ($P$), and molar composition ($c_j$), along with spectroscopic data for each chemical species, including ionization energies, excitation energies, and level degeneracies. Solving this equation yields the key plasma properties and the detailed plasma composition under specified thermodynamic conditions.

\begin{figure}[H]
    \centering
    \includegraphics[width = 0.66\textwidth]{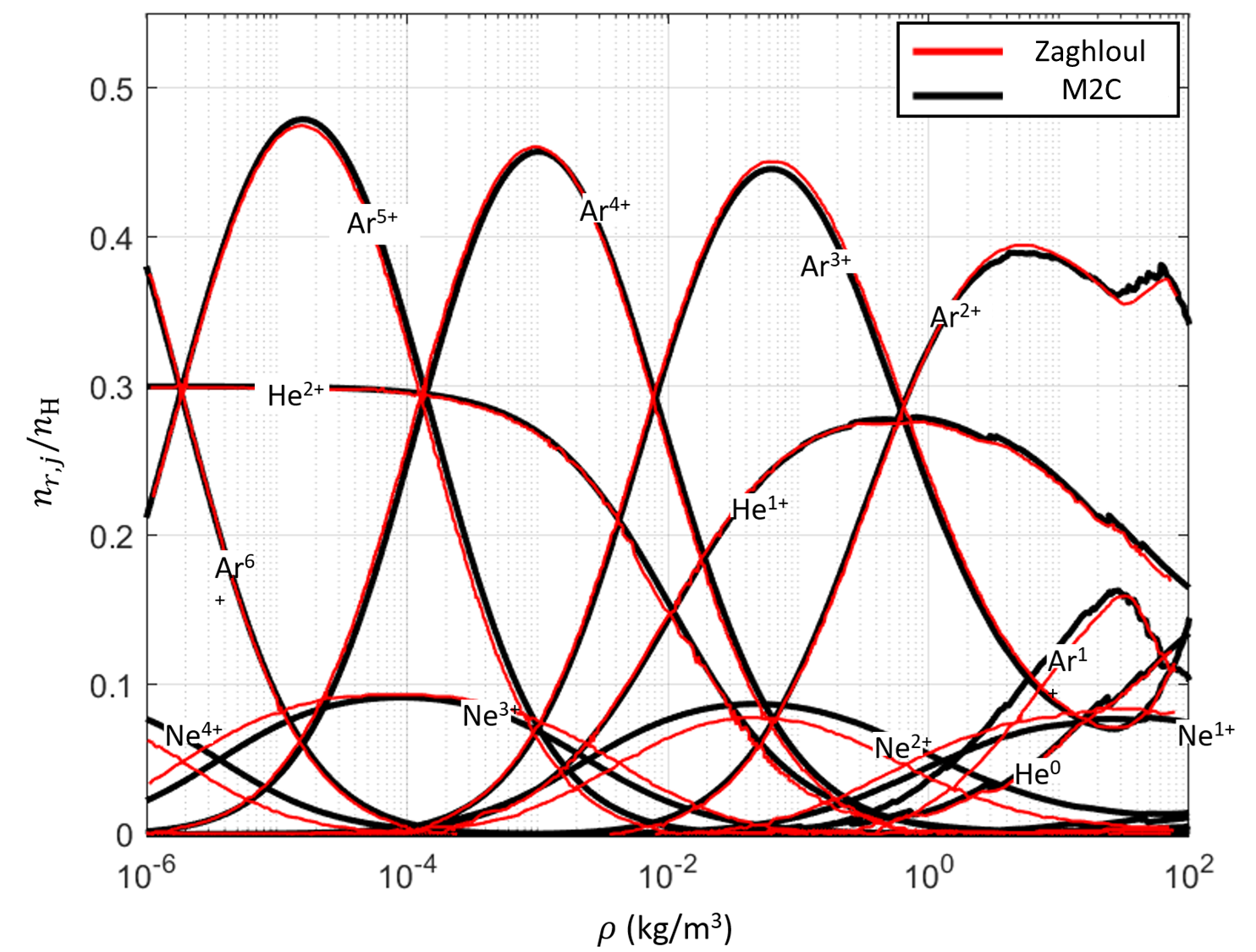}
    \caption{Verification of the non-ideal Saha equation solver implemented in M2C: Dependence of the composition of a $5$ eV non-ideal plasma mixture [He(0.3):Ne(0.6):Ar(0.1)] on density. Reference: Zaghloul~\cite{zaghloul_2004}.}
    \label{fig: non-ideal plasma verification}
\end{figure}

To verify the implementation, the example problem introduce by Zaghloul 
 \cite{zaghloul_2004} is simulated. In this problem,  the plasma composition of a mixture of helium (He), neon (Ne), and argon (Ar) with molar ratios of $0.3:0.1:0.6$ is computed. The plasma is held at a constant temperature of 5eV (approximately 58,000K), while the density varies from $10^{-6}$ to $10^2$kg/m\textsuperscript{3}. As shown in Fig.\ref{fig: non-ideal plasma verification}, the M2C results match reasonably well with the reference solution, particularly for helium and argon ions. Minor discrepancies are observed for neon ions, which may be attributed to numerical errors or algorithmic approximations in either M2C of the reference result.

 For ease of  reproducibility, the ionization module has been packaged as a standalone solver, available at \url{www.github.com/kevinwgy/saha}. Input files and sample output data for this test case are provided in the sub-directory \texttt{Tests/Test1\_HeNeAr} within the \texttt{saha} solver repository.

\subsection{1D multi-material hypervelocity impact analysis}
\label{sec:VandV_hypervelocity}

\begin{figure}
\centering\includegraphics[width= 1\linewidth]{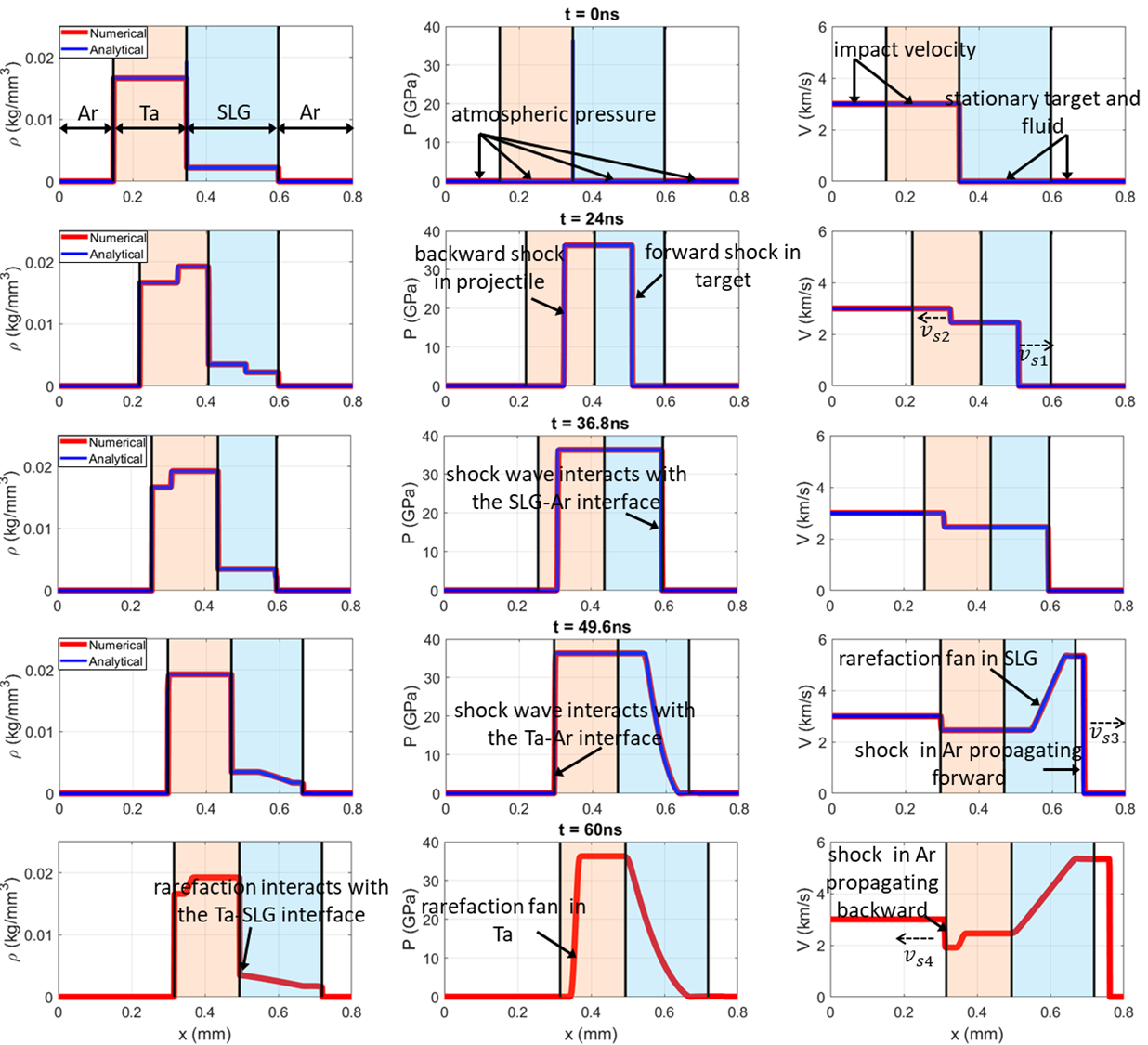}
\caption{1D hypervelocity impact of a tantalum projectile ($V_0 = 3~\text{km/s}$) on a soda lime glass (SLG) target in an atmospheric argon environment.}
\label{fig:1D_HVI}
\end{figure}

This example considers a high-speed planar impact  involving three materials for the projectile, the target, and the ambient fluid (gas). The first row of images in Fig.~\ref{fig:1D_HVI} illustrates the problem setup. The one-dimensional computational domain spans $0 \leq x \leq 1\text{mm}$ along the projectile’s direction of motion and is partitioned into four material subdomains. From left to right, these consist of argon (Ar) gas behind the projectile, the projectile made of tantalum (Ta), the target composed of soda lime glass (SLG), and another region of argon gas behind the target. Further details on the material models and the verification study are available in Ref.~\cite{islam2023fluid}. Replication files for this simulation can be found in the directory
\texttt{Tests/HVImpactShafquatIslam/1D\_HypervelocityImpact}.


The initial conditions for density, velocity, and pressure are specified as follows:

\begin{equation}
\rho(x,0) = 
\begin{cases}
1.78 \times 10^{-3} ~\text{g}/\text{cm}^3 & 0 ~\text{mm} <x \leq 0.15 ~\text{mm},\\
16.65 \text{g}/~\text{cm}^3 & 0.15 ~\text{mm} <x \leq 0.35 ~\text{mm},\\
2.204 \text{g}/~\text{cm}^3 & 0.35 ~\text{mm} <x \leq 0.60 ~\text{mm},\\
1.78 \times 10^{-3} ~\text{g}/\text{cm}^3 & 0.60 ~\text{mm} <x \leq 1 ~\text{mm},\\
\end{cases} 
\end{equation}

\begin{equation}
u(x,0)= 
\begin{cases}
3 ~\text{km}/\text{s} & 0 ~\text{mm} <x \leq 0.35 ~\text{mm},\\
0 ~\text{km}/\text{s} & 0.35 ~\text{mm} <x \leq 1 ~\text{mm},\\
\end{cases}
\end{equation}

and

\begin{equation}
    p(x,0) = 100 ~\text{kPa} \quad 0~\text{mm} \leq x \leq 1~\text{mm}.
\end{equation}

The simulation starts at the moment of collision and proceeds until $t = 60~\text{ns}$. Figure~\ref{fig:1D_HVI} displays the density, velocity, and pressure fields at five time instances:
$t = 0$, $24$, $36.8$, $49.6$, and $60$~ns. At each of these time steps, the numerical solution shows close agreement with the analytical solution wherever available. The initial state is shown in the first row of Fig.\ref{fig:1D_HVI}. The white region denotes argon gas, the orange region represents the Ta projectile, and the blue region indicates the SLG target. The projectile and its trailing gas are initialized at an impact velocity of $3\text{km/s}$, while the target and the forward argon are stationary.

Upon impact, two shock waves are generated at the Ta-SLG interface: a forward-propagating shock in the SLG and a backward-propagating shock in the Ta. These are visible in the snapshot at $t = 24~\text{ns}$. As these shock waves traverse their respective materials, they eventually reach the solid-gas interfaces, triggering reflected rarefaction fans due to the large impedance mismatch. The SLG-Ar interface is encountered first, producing a rarefaction fan that propagates back into the target. This interaction is represented as a Riemann problem and is captured in the $t = 36.8~\text{ns}$ snapshot. At this point, the mass density across the interface changes by over three orders of magnitude, from $3.47~\text{g/cm}^3$ to $1.78 \times 10^{-3}~\text{g/cm}^3$. A similar phenomenon occurs when the shock in the Ta projectile reaches the Ta–Ar interface (shown at $t = 49.6~\text{ns}$), resulting in a rarefaction wave traveling back into the projectile and a transmitted shock propagating into the gas. Up to $t = 60~\text{ns}$, all wave-interface interactions can be modeled as a sequence of one-dimensional Riemann problems, each admitting an exact solution. However, once the rarefaction fan in the SLG reaches the Ta-SLG interface, exact solutions are no longer available. Beyond this point, continued evolution of the impact dynamics can only be simulated numerically.

\section{Illustrative examples}
\label{sec:example}

\subsection{Fluid-structure coupled simulation of explosion containment}
\label{sec:exmp_FSI}

We demonstrate the FSI simulation capabilities of M2C using a test case involving a lightweight explosion containment chamber interacting with fluid flow generated by an internal detonation. The confined geometry leads to complex fluid dynamics, including shock reflections and interactions. The high fluid pressure causes the chamber wall to undergo large, plastic deformations. M2C is used to analyze the fluid dynamics inside the chamber. It is coupled with the Aero-S~\cite{aeros} solver, which analyzes the structural response.

Following conventional pressure vessel design, the chamber consists of a cylindrical midsection with two ellipsoidal end caps. The inner diameter of the cylindrical section is $320~\text{mm}$. The total length of the chamber, including end caps, is $360~\text{mm}$ measured at the inner surface. The chamber wall has a uniform thickness of $5~\text{mm}$ and is made of steel with density $\rho = 7.9\times 10^{-3}~\text{g}/\text{mm}^3$, Young's modulus $E = 210~\text{GPa}$, Poisson's ratio $\nu = 0.3$, and yield strength $\sigma_\text{Y}=355~\text{MPa}$. The material is modeled using J2 plasticity and is assumed to be perfectly plastic beyond its elastic limit. The fluid inside the chamber is assumed to be air and initialized using results of the spherical detonation simulation presented in Sec.~\ref{sec:VandV_expansion}.

The input files for this simulation are available in the directory \texttt{Tests/ExplosionChamberNarkhede}. Additional details regarding the geometry, material properties, and computational setup can be found in~\cite{narkhede2025fluid}.

\begin{figure}[H]
    \centering
    \includegraphics[width=\linewidth]{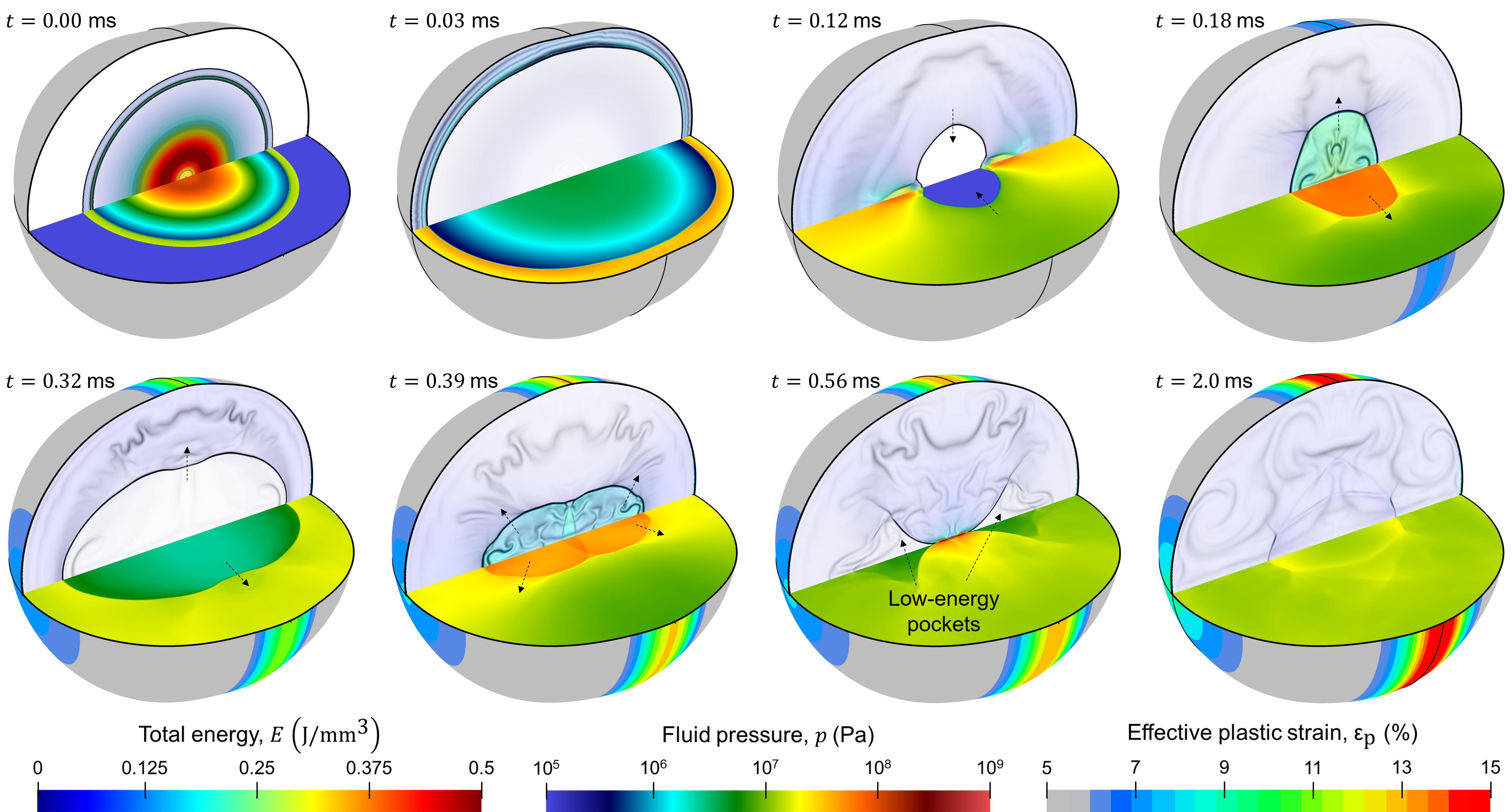}
    \caption{Solution snapshots from the explosion containment simulation. Visualizations show the temporal evolution of total energy per unit volume and fluid pressure, as well as the accumulated plastic strain in the structure.}
    \label{fig:fsi_overview}
\end{figure}

Figure~\ref{fig:fsi_overview} illustrates the fluid and structural results obtained from the simulation. 
 The repeated loading from high-pressure shock waves results in the accumulation of significant plastic strains in the structure.  Due to geometric and stiffness differences between the cylindrical midsection and the ellipsoidal end caps, the deformation is non-uniform. This non-uniformity influences the fluid dynamics, as evidenced by the distortion of the reflected shock fronts, which progressively lose their initial spherical symmetry. These results emphasize the importance of fully coupled simulations. Decoupled or simplified approaches would fail to capture the mutual influence between the structural deformation and internal fluid dynamics, potentially leading to inaccurate predictions of both structural response and pressure evolution.

\subsection{Cavitation induced by long-pulsed laser}
\label{sec:exmp_laser_cavitation}

To illustrate the laser–fluid interaction and vaporization modeling capabilities of M2C, we simulate the formation and growth of a vapor bubble induced by a Holmium:YAG laser beam. The computational domain consists of a cylindrical volume of liquid water into which the laser beam is delivered through an optical fiber of diameter $0.365~\text{mm}$.  The fiber is represented as a fixed embedded boundary within the domain. At the fiber tip,the laser source is specified using a Gaussian spatial irradiance profile, and its temporal power history is prescribed based on experimental measurements. The laser pulse lasts approximately $150~\mu\text{s}$, with a total energy of $0.2~\text{J}$ and a peak power of about $3~\text{kW}$.

The simulation begins with a uniform liquid phase and no pre-existing bubbles. As laser energy is absorbed by the fluid, localized heating near the fiber tip triggers phase transition, leading to the nucleation and subsequent growth of a vapor bubble. The spatial variation of laser irradiance is modeled as described in  Sec.\ref{sec:laser_rad}, and  energy absorption is implementation via the radiation source term in Eq.~\eqref{eq:NSEquation}. The liquid-to-vapor phase transition follows the model detailed in Sec.~\ref{sec:phase_trans}.  Input files and sample results for this simulation are available in the directory \texttt{Tests/LaserCavitationZhao24}. Further details on the numerical setup and parameter values can be found in~\cite{zhao2024vapour}.

\begin{figure}[H]
 \centering
 \includegraphics[width = 0.99 \textwidth]{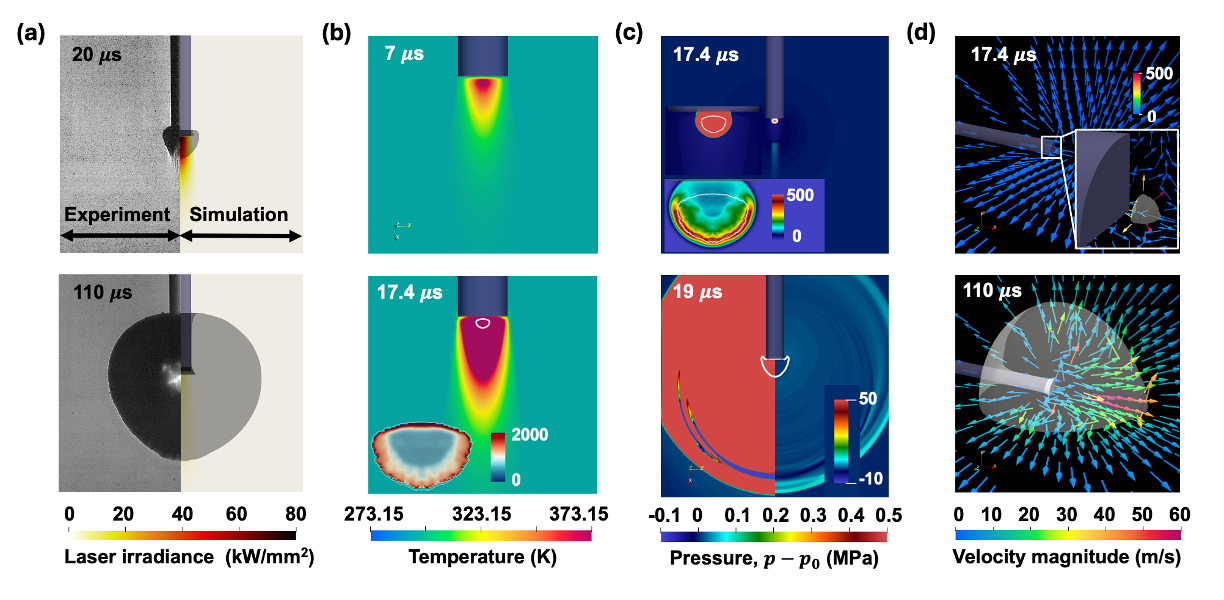}
 \caption{Cavitation bubble generated by long-pulsed Holmium:YAG laser.
(a) Comparison of bubble dynamics from numerical simulation and laboratory experiment at two time instances. (b) Snapshots of the temperature field at $7~\mu\text{s}$ and $17.4~\mu\text{s}$. (c) Pressure field during the early bubble expansion stage, shown at two time instances.
(d) Velocity field at bubble nucleation and at the end of the simulation.}
 \label{fig:Ho_results}
\end{figure}

Figure~\ref{fig:Ho_results} shows the simulated dynamics of a pear-shaped cavitation bubble induced by the laser beam, along with a comparison to experimental images and representative field visualizations of temperature, pressure, and velocity. The simulation predicts the nucleation and asymmetric expansion of the bubble, with size and shape in reasonable agreement with experimental observations. Key physical processes are captured, including localized heating from laser absorption, rapid temperature rise preceding nucleation, high pressure and temperature within the newly formed bubble, and asymmetric flow fields.

\subsection{Hypervelocity impact}
\label{sec:exmp_hypervelocity}

The M2C solver has been applied to simulate high-speed projectile impact impact events, where solid materials experience extreme pressures and temperatures and exhibit fluid-like behavior. As an example, we consider a tantalum rod impacting a soda-lime glass (SLG) target in an argon environment at $5~\text{km}/\text{s}$. This velocity exceeds the speed of sound in both the projectile and the target under ambient conditions, classifying the event as hypervelocity impact. The problem involves strong shock waves in both solids and gas, large structural deformations, and complex interface dynamics.

\begin{figure}[H]
 \centering
 \includegraphics[width = 0.5 \textwidth]{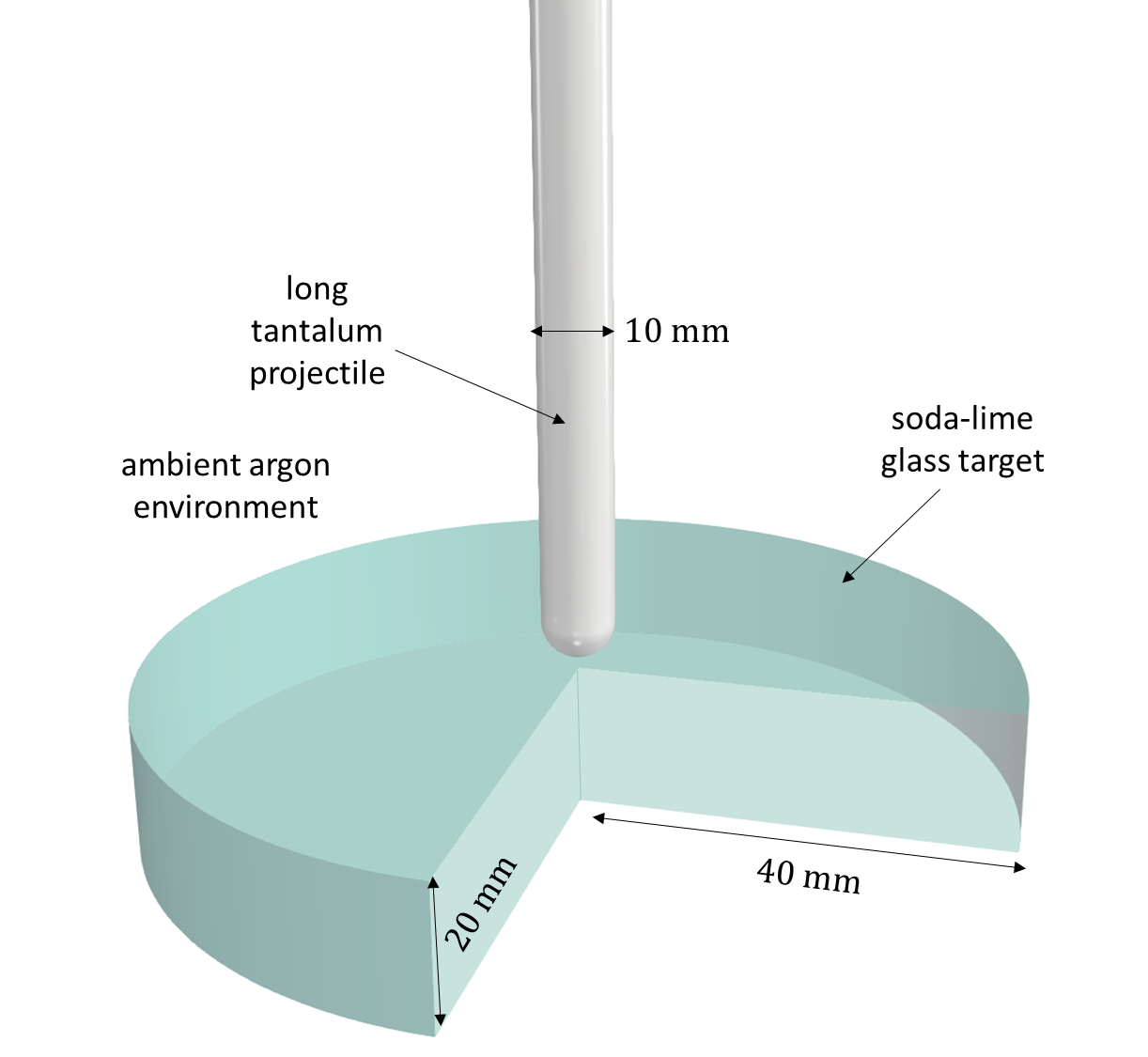}
 \caption{Setup of the hypervelocity impact simulation}
 \label{fig:HVI_setup}
\end{figure}

Figure~\ref{fig:HVI_setup} illustrates the problem setup, including the dimensions of the projectile and target. The computational domain includes  both solid materials (the projectile and the target) as well as the surrounding gas. The boundaries of the projectile and target are tracked using two level set equations that share the same velocity field. The advective mass, momentum, and energy fluxes across material interfaces are computed using the FIVER method. The thermodynamic behavior of tantalum, SLG, and argon is modeled using Mie-Gr\"uneisen, Nobel-Abel stiffened gas, and perfect gas equation of state, respectively. In general, under hypervelocity conditions, the materials may undergo phase transitions, ionization, and chemical reactions. In this simulation, we account for ionization by solving the Saha equation within all three material subdomains. For tantalum and SLG, the generalized (i.e., non-ideal) version of Saha equation is solved, using Ebeling's model to account for the depression of ionization energy. SLG poses an additional complexity as it is a mixture of silica and various metallic silicates. To simplify the ionization modeling, we neglect the bond energies  and treat SLG as a mixture of five pure elements: silicon, oxygen, sodium, calcium, and magnesium. Further details of these models can be found in Ref.~\cite{islam2025ionization}. The input files for this simulation, including all model parameter values, are available in the directory \texttt{Tests/HVImpactShafquatIslam/2D\_axisym\_HVI}.

\begin{figure}[H]
    \centering
    \includegraphics[width = 1\textwidth]{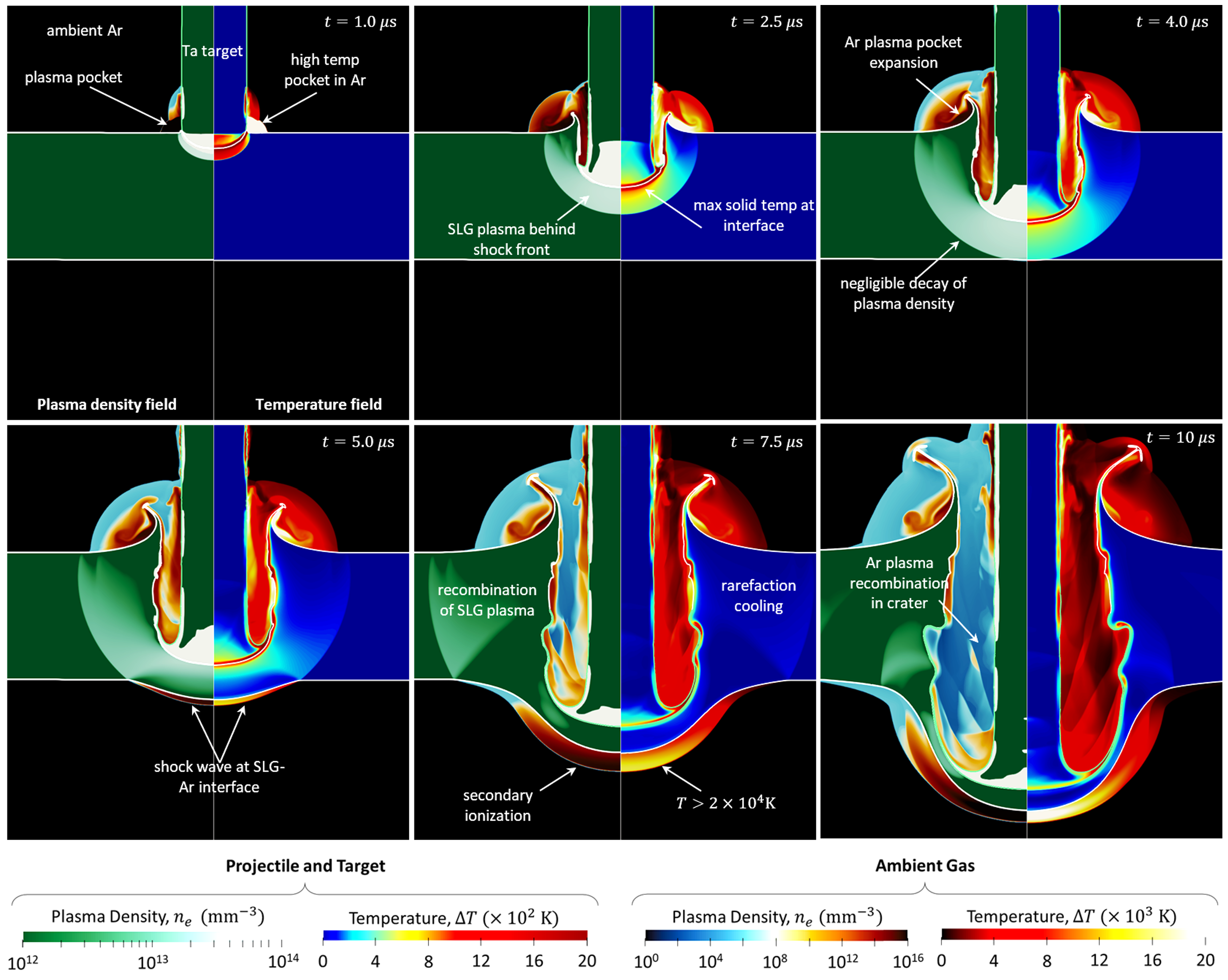}
    \caption{Evolution of temperature and plasma density during a hypervelocity impact event. Impact velocity: $5~\text{km}/\text{s}$}
    \label{fig: Temperature fields}
\end{figure}

Figure~\ref{fig: Temperature fields} presents the temperature and plasma density (i.e., electron number density) results at six time instances during the impact event. Due to the large differences in magnitude across material subdomains, the temperature and plasma density fields are plotted using separate scales for the fluid and the solids. During the impact, the peak temperature in the argon gas reaches values several orders of magnitude higher than those in the projectile and target, even though only a small fraction of the projectile's kinetic energy is transferred to the gas. This pronounced heating is primarily due to the significant mass density disparity between the gas and the solids. As a result, ionization is much more pronounced in the fluid region than in the solid materials. Overall, the simulation captures the propagation of shock waves within all three materials, as well as the large deformation of the projectile and target. A new material interface forms upon contact between the projectile and the target, and the solver successfully handles this interface evolution. The solid-gas interfaces feature density jumps of  three to four orders of magnitude, yet the solver remains robust in tracking and resolving these sharp discontinuities.

\section{Concluding remarks}
\label{sec:conclusion}

M2C (Multiphysics Modeling and Computation) is an open-source software for simulating compressible flow dynamics, including scenarios involving multiple material subdomains and coupled physical processes. It is well-suited for multi-material fluid flows and fluid-structure interaction problems involving shock waves, high pressures, high temperatures, and large structural deformations. 

The design of M2C addresses two challenges: (1) integrating advanced computational geometry algorithms --- such as interface tracking --- and sophisticated material models into computational fluid dynamics, and (2) providing a user-friendly platform for researchers who understand the physics of their problem but lack expertise in numerical methods or high-performance computing. The code adopts a modular, object-oriented architecture and is parallelized using message passing interface (MPI). Simulation inputs are specified through a structured text file. Mesh generation is handled internally; the user only needs to define the computational domain and refinement preferences. Simulation outputs include full-domain solution, sensor data, and plane or line probes. In additional, M2C supports real-time terminal visualization that allows users to monitor simulations as they run. 

The capabilities of M2C are demonstrated in this paper through a series of verification, validation, and application examples. These include academic benchmark problems such as Riemann problems, interface evolution tests, and ionization response. Additional examples address complex multiphysics scenarios, including laser-induced thermal cavitation, bubble dynamics, cavitation erosion, explosion and mitigation, and hypervelocity impact. These test cases illustrate the robustness and versatility of the code.

\section*{Acknowledgment}

The authors gratefully acknowledge support from the National Science Foundation (NSF) under Award No.~CBET-1751487; the Office of Naval Research (ONR) under Award Nos.~N00014-19-1-2102 and N00014-24-1-2509; the Naval Air Warfare Center Aircraft Division under Contract No.~N6833525C0444; the U.S. Department of Transportation (DOT) Pipeline and Hazardous Materials Safety Administration under Contract No.~693JK32250007CAAP; and the National Institutes of Health (NIH) under Award Nos.~2R01-DK052985-26 and R01-DK138972. The development of M2C was inspired and facilitated by these grants, although it was not a deliverable for any of them.

\appendix
\section{Organization of files and sub-directories}
\label{app:files_org}

Table \ref{tab:files} categorizes the files (excluding .cpp files) and directories in M2C into groups for clarity and ease of reference. A description of each file’s purpose can be found at the beginning of the header file.

\begin{small} 
\renewcommand{\arraystretch}{1.2} 

\begin{longtable}{p{2cm}|p{4.2cm}|p{9cm}}
    \caption{Files and folders in M2C. }     \label{tab:files} \\
    \hline
    \multicolumn{2}{l|}{\textbf{Component}} & Files/Folders (Exclude .cpp files) \\
    \hline
    \endfirsthead

    \hline
    \multicolumn{2}{l|}{\textbf{Component}} & Files/Folders (Exclude .cpp files)\\
    \hline
    \endhead

    \hline
    \endfoot

    \multicolumn{2}{l|}{\textbf{Input}} & IoData.h, PrescribedMotionOperator.h, UserDefinedState.h, UserDefinedDynamics.h, UserDefinedForces.h, UserDefinedSolution.h  \\
    \hline
    \multicolumn{2}{l|}{\textbf{Mesh generation}} & GlobalMeshInfo.h, MeshGenerator.h, MeshMatcher.h \\
    \hline
    \multirow{5}{{2.3cm}}{\textbf{Finite-volume solver}} & Equations of state (EOS) & EOSAnalyzer.h, VarFcnANEOSBase.h, VarFcnANEOSEx1.h, VarFcnBase.h, VarFcnDummy.h, VarFcnHomoIncomp.h, VarFcnJWL.h, VarFcnMGExt.h, VarFcnMG.h, VarFcnNASG.h, VarFcnSG.h, VarFcnTillot.h \\
    \cline{2-3}
     & Multi-material fluids & MultiPhaseOperator.h, SpaceInitializer.h, SpaceOperator.h, SpaceVariable.h, TimeIntegrator.h, Main.cpp\\
    \cline{2-3}
    & Exact riemann solver & KDTree.h, ExactRiemannSolverBase.h, ExactRiemannSolverInterfaceJump.h, RiemannSolutions.h \\
    \cline{2-3}
    &Flux computation & FluxFcnBase.h, FluxFcnGenRoe.h, FluxFcnGodunov.h, FluxFcnHLLC.h, FluxFcnLLF.h, Reconstructor.h, SymmetryOperator.h \\
    \cline{2-3}
    & Additional physics  & GravityHandler.h, HeatDiffusionFcn.h, HeatDiffusionOperator.h, ViscoFcn.h, ViscosityOperator.h \\
    \hline
    \multicolumn{2}{l|}{\textbf{Level set}}  & LevelSetOperator.h, LevelSetReinitializer.h\\
    \hline
    \multicolumn{2}{l|}{\textbf{Fluid-structure interaction}}   & CrackingSurface.h, DynamicLoadCalculator.h, EmbeddedBoundaryDataSet.h, EmbeddedBoundaryFormula.h, EmbeddedBoundaryOperator.h, Intersector.h, MultiSurfaceIntersector.h, TimeIntegratorSemilmp.h, TriangulatedSurface.h \\
    \hline
    \multicolumn{2}{l|}{\textbf{Ionization}} & AtomicIonizationData.h, IonizationOperator.h, NonIdealSahaEquationSolver.h, SahaEquationSolver.h \\
    \hline
    \multicolumn{2}{l|}{\textbf{Laser}} & LaserAbsorptionSolver.h \\
    \hline
    \multicolumn{2}{l|}{\textbf{Phase change}} & PhaseTransition.h \\
    \hline
    \multicolumn{2}{l|}{\textbf{Output}} & EnergyIntegrationOutput.h, LagrangianOutput.h, MaterialVolumeOutput.h, Output.h, ParaviewRemoteVisual/, PhaseTransitionOutput.h, PlaneOutput.h, ProbeOutput.h, TerminalVisualization.h \\
    \hline
    \multirow{4}{*}{\textbf{Utilities}} & MPI communications & CommunicationTools.h, CustomCommunicator.h, GhostFluidOperator.h, NeighborCommunicator.h \\
    \cline{2-3}
    & External solver coulping & AerofMessenger.h, AerosMessenger.h, ConcurrentProgramsHandler.h, M2CTwinMessenger.h \\
    \cline{2-3}
    & Computational tools  & ClosestTriangle.h, FloodFill.h, GeoTools/, GhostPoint.h,GradientCalculatorBase.h, GradientCalculatorCentral.h, GradientCalculatorFD3.h, Interpolator.h, LinearOperator.h, LinearSystemSolver.h, MathTools/, parser/, ReferenceMapOperator.h, SmoothingOperator.h, SpecialToolsDriver.h, SteadyStateOperator.h, Utils.h, Vector2D.h, Vector3D.h, Vector5D.h \\
    \cline{2-3}
    & Compilation & cmake/, CMakeLists.txt, COPYING.txt, version.cmake \\
    \hline
    \multicolumn{2}{l|}{\textbf{Tests}} & Tests/, Restart/\\
    \hline
    \multicolumn{2}{l|}{Additional modules not covered in this paper} & HyperelasticityFcn.h, HyperelasticityFcn2DCyl.h, HyperelasticityOperator.h, IncompressibleOperator.h\\
    \hline
\end{longtable}
\end{small}

\section{Equations of State}
\label{app:eos}
M2C supports a variety of equations of state (EOSs) to accommodate different materials and simulation objectives. Users may select an appropriate EOS based on material properties, thermodynamic behavior, and the intended application. Below, we summarize the mathematical formulations of several EOSs implemented in M2C.


\subsection{Noble-Abel stiffened gas}
The Noble-Abel stiffened gas (NASG) EOS~\cite{NobelAbel2016} is implemented in \texttt{VarFcnNASG.h}. It is commonly used for modeling compressible materials subjected to high pressures. It is a generalization of perfect gas (PG), stiffened gas (SG), and the Noble–Abel EOS. The corresponding pressure and temperature equations are presented in Section~\ref{sec:EOS}. M2C also provides a separate implementation of the SG EOS in \texttt{VarFcnSG.h}. Materials obeying the PG or SG models can be defined using either the SG or NASG EOS option.

\subsection{Jones-Wilkins-Lee}
The Jones-Wilkins-Lee (JWL) equation of state is widely used to model gaseous detonation products~\cite{menikoff2015jwl}. It is implemented in \texttt{VarFcnJWL.h}. The pressure equation has the expression
\begin{equation}
p=\omega \rho e + A_{1}\left(1-\frac{\omega \rho}{R_{1} \rho_{0}}\right)  \exp \Big(\frac{R_{1} \rho_{0}}{\rho}\Big)+A_{2}\left(1-\frac{\omega \rho}{R_{2} \rho_{0}}\right) \exp \Big(-\frac{R_{2} \rho_{0}}{\rho} \Big),
\label{eq:JWL_EOS}
\end{equation}
where $\rho_0$, $\omega$, $A_1$, $A_2$, $R_1$, and $R_2$ are constant parameters. The temperature law is given by
\begin{equation}
T = \dfrac{1}{c_{v}} \Big( e - \dfrac{A_1}{\rho_0 R_1} \exp \big( - \dfrac{R_1 \rho_0}{\rho}\big)  -  \dfrac{A_2}{\rho_0 R_2} \exp \big( - \dfrac{R_2 \rho_0}{\rho}\big)  \Big),
\label{eq:JWL_Tlaw}
\end{equation}
where $c_v$ is assumed to be constant. 

\subsection{Mie-Gr\"{u}neisen}
The Mie-Gr\"{u}neisen (MG) equation of state is implemented in \texttt{VarFcnMGExt.h}. It is often used to model solids and liquids under high pressures. The formulation presented in~\cite{Robinson2019mie} is adopted, which accounts for material behavior under both compression and tension. In both regimes, the pressure and temperature equations follow the general expressions
\begin{align}
p(\eta,e) &= p_R(\eta) + \rho_0\Gamma_0\big(e - e_R(\eta)\big),\\
e(\eta,T) &= e_R(\eta) + c_v\big(T - T_R(\eta)\big),
\end{align}
where $\eta = 1 - \rho_0/\rho$ denotes the volumetric strain relative to a reference density $\rho_0$, and $\Gamma_0$ is the Gr\"{u}neisen parameter at the reference state. $c_v$ is the specific heat capacity at constant volume. $p_R(\eta)$, $e_R(\eta)$, and $T_R(\eta)$ are defined separately for compression and tension.

If the material is in compression (i.e., $\eta>0$),
\begin{align}
p_R(\eta) &=  \dfrac{\rho_0 c_0^2 \eta}{(1-s \eta)^2},\\
e_R(\eta) &=  \dfrac{p_H(\eta)\eta}{2\rho_0} + e_0,\\
T_R(\eta) &=  e^{\Gamma_0 \eta}\left(T_0 + \dfrac{c_0^2 s}{c_v} \int_0^\eta \dfrac{e^{\Gamma_0 z}z^2}{(1-sz)^3} dz \right).
\end{align}
where $c_0$ denotes the speed of sound at the reference state, and $s$ the slope of the assumed linear relation between shock speed and particle velocity in shock compression experiment. $e_0$ is the internal energy at the reference state, which can be set to $0$ in many cases.

If the material is in tension with relatively small strain (i.e., $\eta_{\text{min}} \leq \eta < 0$, where $\eta_{\text{min}}$ is a user-specified threshold),
\begin{align}
p_R(\eta) &= K_0 \eta,\\
e_R(\eta) &= \dfrac{K_0 \eta^2}{2\rho_0} + e_0,\\
T_R(\eta) &= T_0 e^{\Gamma_0\eta},
\end{align}
with $K_0 = \rho_0c_0^2$ ensuring the continuity of $p_R$ and its first derivative across $\eta = 0$. $T_0$ is the temperature at the reference state. 

For strains exceeding the tensile limit (i.e., $\eta < \eta_{\text{min}}$),
\begin{align}
p_R(\eta) &= p_{\text{min}},\\
e_R(\eta) &= \dfrac{K_0\eta^2_{\text{min}}}{2\rho_0} + e_0 + \dfrac{p_{\text{min}}}{\rho_0}(\eta - \eta_{\text{min}}),\\
T_R(\eta) &= T_0 e^{\Gamma_0 \eta},
\end{align}
with $p_\text{min} = K_0\eta_{\text{min}}$ represents the minimum pressure (i.e., maximum tensile stress) permitted for the material.

\subsection{Tillotson}
\begin{figure}[h!]
\includegraphics[width=0.5\linewidth] {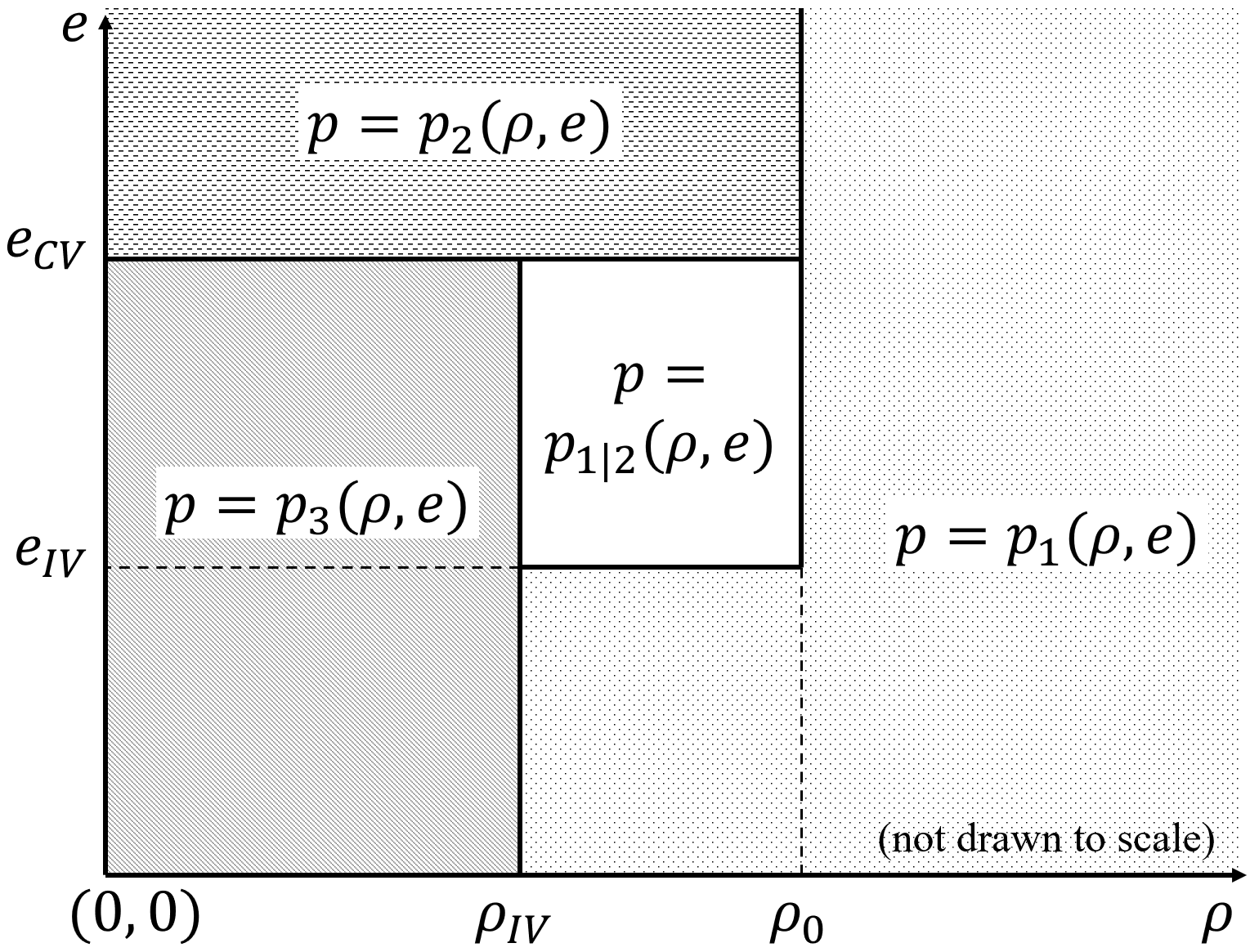}
\centering
\caption{Illustration of the extended Tillotson EOS implemented in M2C.}
\label{fig:Tillotson_cases}
\end{figure}

The Tillotson EOS is implemented in \texttt{VarFcnTillot.h}. Originally presented in \cite{tillotson1962metallic}, it is often used to model solid materials under high pressures and temperatures. M2C adopts the extended formulation presented in~\cite{brundage2013implementation}, which partitions the $\rho$-$e$ space into four distinct regions (Fig.~\ref{fig:Tillotson_cases}). It consists of three basic functions and accounts for phase transition by defining a fourth ``mixing region'' between the vapor and condensed phases, where the pressure is a linear blend of the two states. Specifically, the pressure equation is defined by
\begin{equation}
p =
\begin{cases}
\begin{aligned}
p_1 &= (a + b\chi ) \rho e + A (\frac{\rho}{\rho_0} - 1) + B (\frac{\rho}{\rho_0} - 1)^2, \\
& \phantom{\makebox[5em]{}} \text{if } \rho \geq \rho_0, e \geq 0 \text{ or } \rho_{IV} \leq \rho < \rho_0, 0 \leq e \leq e_{IV}, \\
p_2 &= a\rho e + \big(b \rho e \chi + A (\frac{\rho}{\rho_0} - 1) \exp(\beta -\frac{\beta \rho_0}{\rho})\big) \exp\big(-\alpha (\frac{\rho_0}{\rho}-1)^2 \big), \\
& \phantom{\makebox[5em]{}} \text{if } \rho < \rho_0, e \geq e_{CV}, \\
p_3 &= (a + b\chi) \rho e + A (\frac{\rho}{\rho_0} - 1), \\
& \phantom{\makebox[5em]{}} \text{if } \rho < \rho_{IV}, 0 \leq e < e_{CV}, \\
p_{1|2} &= \frac{(e_{CV} - e)p_1 + (e - e_{IV})p_2}{e_{CV} - e_{IV}}, \\
& \phantom{\makebox[5em]{}} \text{if } \rho_{IV} \leq \rho < \rho_0, e_{IV} < e < e_{CV}.
\end{aligned}
\end{cases}
\label{eq:Tillo_EOS}
\end{equation}
with $\chi = (e_0 \rho^2)/(e\rho_0^2 + e_0 \rho^2)$. Here, $a$, $b$, $A$, $B$, $\alpha$, and $\beta$ are constant model parameters. $\rho_0$ and $e_0$ define a reference state, which must be in a condensed phase. $\rho_{IV}$ and $e_{IV}$ are the density and internal energy per unit mass that correspond to ``incipient vaporization'', while $e_{CV}$ is the internal energy per unit mass corresponding to ``complete vaporization''. The temperature is calculated by
\begin{equation}
T = T_0 + \frac{e - e_c}{c_v},
\label{eq:Tillo_Tlaw}
\end{equation}
where $T_0$ is the temperature corresponding to $e = e_c(\rho_0)$. $e_c(\rho)$ is obtained by integrating
\begin{equation}
\dfrac{de_c}{d\rho} = \dfrac{p(\rho,e_c)}{\rho^2}
\end{equation}
with initial condition $\rho=\rho_0$, $e_c =0$.

\subsection{ANEOS: Physics-based EOS}
ANEOS (ANalytic EOS) is a class of EOS that models the Helmholtz free energy,
\begin{equation}
F(\rho,T) = e - Ts,
\end{equation}
where $F$ is the specific Helmholtz free energy, and $s$ the specific entropy. A general model form is
\begin{equation}
F(\rho, T) = F_{\text{cold}}(\rho) + F_{\text{thermal}}(\rho,T) + F_{\text{electronic}}(\rho,T),
\label{eq:ANEOS_helm}
\end{equation}
where $F_{\text{cold}}(\rho)$ accounts for the atomic interactions that do not depend on temperature, such as the interatomic potential. $F_{\text{thermal}}$ contains the temperature-dependent part of the interatomic forces. $F_{\text{electronic}}$ accounts for energies related to ionization of atoms and is only important at very high temperatures and low densities~\cite{melosh2007hydrocode}.

Using the first law of thermodynamics, the differential of $F$ can be written as
\begin{equation}
dF = -sdT + \dfrac{p}{\rho^2}d\rho,
\label{eq:helmholtz_differential}
\end{equation}
which can be used to derive several key thermodynamic variables. For example,
\begin{align}
s &=  -\dfrac{\partial F}{\partial T}\Big|_{\rho}~~~~\text{(entropy),}\label{eq:ANEOS_entropy_def}\\
p &= \rho^2\dfrac{\partial F}{\partial \rho}\Big|_{T}~~~~\text{(pressure),}\\
e &= F + Ts~~~~\text{(internal energy),}\label{eq:ANEOS_e_F_Ts}
\end{align}

M2C implements a specific ANEOS model described in~\cite{sanchez2021inelastic}, available in the file \texttt{VarFcnANEOSEx1.h}. In this model, $e$ and $F$ are defined as
\begin{align}
    e(\rho,T) = e_c(\rho) + e_l(\rho,T) + \Delta e,  \label{eq:ANEOS_e} \\
    F(\rho,T) = e_c(\rho) + F_l(\rho,T) + \Delta e,  \label{eq:ANEOS_F}
\end{align}
where $e_c$ and $e_l$ represent the contributions of interatomic potential energy and thermal vibrations. $\Delta e$ represents an energy shift constant. $e_c$ is modeled using the Birch-Murnaghan equation of state:
\begin{equation}
e_c(\rho) = \dfrac{9b_0}{8r_0} \Big( \big(\dfrac{\rho}{r_0}\big)^{\frac{2}{3}} - 1\Big)^2 \Big[ \dfrac{1}{2}\Big(\big(\dfrac{\rho}{r_0}\big)^{\frac{2}{3}} - 1\Big)(b_0'-4) + 1\Big].
\end{equation}

$e_l$ and $F_l$ are given by the Debye model,
\begin{align}
e_l(\rho,T) = & \dfrac{R}{w}\Big[\dfrac{9}{8} \Theta + \dfrac{9T^4}{\Theta^3}\int_0^{(\Theta/T)} \dfrac{y^3}{e^y - 1} dy \Big],\\
F_l(\rho,T) = & \dfrac{R}{w}\Big[\dfrac{9}{8}\Theta + 3T \ln\Big(1 -\exp \big(-\frac{\Theta}{T} \big)\Big) - \dfrac{3T^4}{\Theta^3}\int_0^{(\Theta/T)} \dfrac{y^3}{e^y - 1} dy \Big],
\label{eq:Debye_model}
\end{align}
with $\Theta= T_0(\rho/\rho_0)^{\Gamma_0}$. Here, $b_0$, $b_0'$, $T_0$, $\Gamma_0$, $w$, and $R$ are constant model parameters, while $r_0$ and $\rho_0$ are the reference density at $0$ Kelvin and ambient conditions, respectively.

This ANEOS model is a complete EOS: no separate temperature equation is needed. The temperature $T$ is determined by numerically solving Eq.~\eqref{eq:ANEOS_e} for given $\rho$ and $e$. Then, pressure $p$ is evaluated via
\begin{equation}
p(\rho,e) = \rho^2 \dfrac{\partial F}{\partial \rho}\Big|_T\nonumber = \rho^2 \Big( \dfrac{d e_c(\rho)}{d\rho} + \dfrac{\partial F_l(\rho,T)}{\partial \rho} \Big).
\label{eq:ANEOS}
\end{equation}

In M2C, many EOS-related functions take $\rho$ and $e$ as input arguments, instead of $T$. To avoid repeatedly solving the temperature equation with the same input, M2C maintains and continuously updates an array of recently computed $(\rho,e,T)$ tuples. Before invoking the temperature solver, it first searches this array for a match. If found, the corresponding stored value is reused, avoiding redundant computation.

\section{Numerical flux functions}
\label{app:flux_func}

M2C  implements several numerical flux functions, including local Lax-Friedrichs~\cite{rusanov1970difference}, Roe-Pike~\cite{hu2009hllc}, and HLLC~\cite{toro2013riemann}. Their source codes are in \texttt{FluxFcnLLF.h}, \texttt{FluxFcnGenRoe.h} and \texttt{FluxFcnHLLC.h}, respectively.  An outline of each method is provided below.

\subsection{Local Lax-Friedrichs flux}
 Let $\bm{W}_{ij}$ and $\bm{W}_{ji}$ denote the reconstructed state variables on the two sides of $\partial C_{ij}$, the interface between control volumes $C_i$ and $C_j$. The local Lax-Friedrichs flux function, implemented in \texttt{FluxFcnLLF.h}, is given by
\begin{equation}
    \Phi(\bm{W}_{ij}, \bm{W}_{ji}, \bm{n}_{ij}) = \dfrac{1}{2} (\mathcal{F}(\bm{W}_{ij}) \cdot \bm{n}_{ij} + \mathcal{F}(\bm{W}_{ji}) \cdot \bm{n}_{ij}) - \dfrac{1}{2} \lambda_{max} (\bm{W}_{ji} - \bm{W}_{ij}),
\label{eq:Lax-Friedrichs}
\end{equation}
where $\lambda_{max}$ is the largest eigenvalue (in magnitude) of the Jacobian matrix of the flux in the normal direction (i.e., the fastest wave speed), i.e.,
\begin{equation}
    \lambda_{\max} = \max \left( |u_{ij}| + |c_{ij}|,\ |u_{ji}| + |c_{ji}| \right),
\end{equation}
with 
\begin{equation}
    u_{ab}= \bm{V}_{ab} \cdot \bm{n}_{ij},~ab \in \{ij, ji\}
\end{equation}
being the normal velocity components and $c$ the speed of sound.

\subsection{Flux Jacobians}
The Roe-Pike flux requires the derivatives of the physical flux vectors with respect to the conservative state variables $\bm{W}$. For a general EOS of the form $p = p(\rho,e)$,
\begin{equation}
\bm{A} \equiv \dfrac{\partial (\mathcal{F}(\bm{W})\cdot\bm{n})}{\partial \bm{W}} = \begin{bmatrix}
0 & \bm{n}^T & 0\\
-u_n\bm{V} + \Big(p_\rho+\big(\dfrac{1}{2}|\bm{V}|^2-e\big)\Gamma\Big)\bm{n}  &  u_n\mathbb{I}+\bm{V}\bm{n}^T-\Gamma\bm{n}\bm{V}^T & \Gamma\bm{n}\\
u_n\Big(-H+p_\rho+\big(\dfrac{1}{2}|\bm{V}|^2 - e\big)\Gamma\Big) & -u_n\Gamma\bm{V}^T+H\bm{n}^T & u_n(\Gamma+1)
\end{bmatrix},
\label{eq:flux_Jac}
\end{equation}
where $\bm{n}$ denotes the unit normal vector of the flux direction. Because M2C utilizes Cartesian grids, $\bm{n}$ is $[1,0,0]^T$, $[0,1,0]^T$, or $[0,0,1]^T$ for the $x$, $y$, and $z$ directions, respectively. $u_n = \bm{V}\cdot \bm{n}$ is the normal velocity component.  $\displaystyle H = E + \dfrac{p}{\rho}$ denotes the specific total enthalpy. $\displaystyle p_\rho = \dfrac{\partial p(\rho,e)}{\partial \rho}$, and $\Gamma = \dfrac{1}{\rho}\dfrac{\partial p(\rho,e)}{\partial e}$ is the Gr\"uneisen parameter.
\vspace{2mm}

For example, in the $x$ direction, the physical flux function and its Jacobian matrix are given by
\begin{equation}
\bm{f}(\bm{W}) = \begin{bmatrix}
\rho u \\ \rho u^2 + p \\ \rho u v \\ \rho u w \\\rho H u
\end{bmatrix},
\end{equation}
\begin{equation}
\bm{A} = \dfrac{\partial\bm{f}}{\partial \bm{W}}
=
\begin{bmatrix}
0 & 1 & 0 & 0 & 0\\[3ex]
-u^2 + p_\rho + \big(\dfrac{1}{2}|\bm{V}|^2 - e\big)\Gamma 
& u(2-\Gamma) & -v \Gamma & -w \Gamma & \Gamma \\[3ex]
-uv & v & u & 0 & 0\\[3ex]
-uw & w & 0 & u & 0\\[3ex]
u\Big(-H + p_\rho + \big(\dfrac{1}{2}|\bm{V}|^2-e\big)\Gamma\Big)
& H - \Gamma u^2 & -uv\Gamma & -uw\Gamma  & u(\Gamma + 1)
\end{bmatrix}.
\end{equation}

\vspace{4mm}
The eigen decomposition of $\bm{A}$ in \eqref{eq:flux_Jac} is
\begin{equation}
\bm{A} = \bm{R}\bm{\Lambda}\bm{R}^{-1},
\end{equation}
with
\begin{equation}
\Lambda = \begin{bmatrix}
u_n - c & \bm{0}^T & 0\\[3ex]
\bm{0} &  u_n\mathbb{I}  &  \bm{0} \\[3ex]
0 & \bm{0}^T & u_n+c
\end{bmatrix},
\end{equation}
\begin{equation}
R = \begin{bmatrix}
1               & \bm{n}^T  & 1 \\  
\bm{u}-c\bm{n}  &      \mathbb{I}+\bm{V}\bm{n}^T-\bm{n}\bm{n}^T     & \bm{V}+c\bm{n}\\
H - u_n c       &    \bm{V}^T + (e+\dfrac{1}{2}|\bm{V}|^2 - \dfrac{p_\rho}{\Gamma} - u_n)\bm{n}^T & H+u_n c\\
\end{bmatrix},
\end{equation}
\begin{equation}
R^{-1} = \begin{bmatrix}
\dfrac{1}{2}-\dfrac{\beta}{2}+\dfrac{u_n}{2c}  &  -\dfrac{\alpha}{2}\bm{V}^T -\dfrac{1}{2c}\bm{n}^T &  \dfrac{\alpha}{2}\\[3ex]
-\bm{V}+(\beta+u_n)\bm{n}
& \mathbb{I} + \alpha \bm{n}\bm{V}^T - \bm{n}\bm{n}^T
& -\alpha\bm{n}\\[3ex]
\dfrac{1}{2}-\dfrac{\beta}{2}-\dfrac{u_n}{2c}  &  -\dfrac{\alpha}{2}\bm{V}^T + \dfrac{1}{2c}\bm{n}^T  & \dfrac{\alpha}{2}
\end{bmatrix},
\end{equation}
where $\alpha = \rho\Gamma/(p\Gamma + \rho p_\rho)$, $\beta = (H-|V|^2)\alpha$, and
\begin{equation}
c = \sqrt{\dfrac{\partial p(\rho,e)}{\partial \rho} + \Gamma \dfrac{p}{\rho}} 
\end{equation}
is the speed of sound.

The above formulas for flux Jacobians and their eigen decompositions are implemented in \texttt{FluxFcnBase.h}.

\subsection{Roe-Pike flux}
The original Roe flux was developed for the perfect gas EOS. The Roe-Pike flux implemented in \texttt{FluxFcnGenRoe.h} is a generalization that can be used with any convex EOS. It can be written as
\begin{equation}
    \Phi(\bm{W}_{ij}, \bm{W}_{ji}, \bm{n}_{ij}) = \dfrac{1}{2} (\mathcal{F}(\bm{W}_{ij}) \cdot \bm{n}_{ij} + \mathcal{F}(\bm{W}_{ji}) \cdot \bm{n}_{ij}) - \dfrac{1}{2} \sum_{p=1}^5 |\hat{\lambda}_p| \hat{\alpha}_p \hat{\bm{r}}_p,
\label{eq:Roe-Pike}
\end{equation}
where $\hat{\lambda}_p$ is the $p$-th eigenvalue of the Roe-averaged Jacobian matrix in the normal direction, $\hat{\bm{r}}_p$ is the corresponding eigenvector, and $\hat{\alpha}_p$ is coefficients obtained by projecting the jump $\bm{W}_{ji} - \bm{W}_{ij}$ onto the eigenvectors.

\subsection{HLLC flux}
The Harten-Lax-van Leer-Contact (HLLC) flux is implemented in \texttt{FluxFcnHLLC.h}. HLLC requires estimating the minimum and maximum wave speeds at the
control volume interface, denoted by $S_{ij}$ and $S_{ji}$. It assumes that between these two waves, there is a contact discontinuity across which velocity and pressure are continuous.  The HLLC flux function is defined as
\begin{equation}
    \Phi(\bm{W}_{ij}, \bm{W}_{ji}, \bm{n}_{ij}) = 
    \begin{cases}
        \mathcal{F}(\bm{W}_{ij}) \cdot \bm{n}_{ij}, & \text{if } 0 \le S_{ij}, \\
        \mathcal{F}_{*}^{ij} \cdot \bm{n}_{ij}, & \text{if } S_{ij} \le 0 \le S_*, \\
        \mathcal{F}_{*}^{ji} \cdot \bm{n}_{ij}, & \text{if } S_* \le 0 \le S_{ji}, \\
        \mathcal{F}(\bm{W}_{ji}) \cdot \bm{n}_{ij}, & \text{if } 0 \ge S_{ji},
    \end{cases}
\end{equation}
where $S_*$ is the speed of the contact wave. The intermediate fluxes are computed as
\begin{equation}
    \mathcal{F}_{*}^{ab} = \mathcal{F}(\bm{W}_{ab}) + S_{ab} \left( \bm{W}_{*}^{ab} - \bm{W}_{ab} \right), \quad ab \in \{ij, ji\},
\end{equation}
where the intermediate state $\bm{W}_{*}^{ab}$ is given by
\begin{equation}
    \bm{W}_{*}^{ab} = \rho_{ab} \frac{S_{ab} - u_{ab}}{S_{ab} - S_*} 
    \begin{bmatrix}
        1 \\
        \bm{V}_* \\
        E_*
    \end{bmatrix}.
\end{equation}
The normal velocity and pressure in the star region satisfy
\begin{equation}
    \bm{V}_* \cdot \bm{n}_{ij} = S_*, \quad
    p_* = p_{ab} + \rho_{ab} (S_{ab} - u_{ab})(S_* - u_{ab}).
\end{equation}

The wave speeds can be estimated by
\begin{align}
    S_{ij} &= \min(u_{ij} - c_{ij}, u_{ji} - c_{ji}), \\
    S_{ji} &= \max(u_{ij} + c_{ij}, u_{ji} + c_{ji}), \\
    S_* &= \frac{p_{ji} - p_{ij} + \rho_{ij} u_{ij} (S_{ij} - u_{ij}) - \rho_{ji} u_{ji} (S_{ji} - u_{ji})}{\rho_{ij}(S_{ij} - u_{ij}) - \rho_{ji}(S_{ji} - u_{ji})}.
\end{align}

\section{Slope limiters}
\label{app:limiter}

M2C provides several slope limiter options, including the Monotonized Central (MC) limiter, the Van Albada limiter, and a modified Van Albada limiter. These limiters are implemented in \texttt{Reconstructor.h/cpp}, following the approach described in~\cite{zeng2016}. As an example, the limiter expressions in the $x$ direction are shown below. The subscript $ijk$ denotes the cell index, and $\Delta x_i$ refers to the local control volume width along the $x$ direction.

The Monotonized Central-difference (MC) limiter is define as
\begin{equation}
\phi_{lim}(\theta) = \max\Big(0,~\min(\alpha\theta,~\frac{B}{A+1}(\theta+1),~\alpha)\Big),
\label{eq:theta}
\end{equation}
where
\begin{equation}
\theta = \frac{w_{ijk}-w_{i-1,jk}}{w_{i+1,jk}-w_{ijk}}, \quad
A = \frac{\Delta x_{i-1} + \Delta x_{i}}{\Delta x_i + \Delta x_{i+1}},\quad
B = \frac{2 \Delta x_i}{\Delta x_i + \Delta x_{i+1}}.
\label{eq:AB}
\end{equation}
The parameter $\alpha$ is set to $1.2$ by default. Setting $\alpha = 0$ gives constant reconstruction. $\alpha$ can be tuned between $1$ and $2$, with larger values producing less dissipation~\cite{kurganov2002}.

The Van Albada limiter is define as
\begin{equation}
\phi_{lim}(\theta) = \frac{B (\theta^k + \theta)}{\theta^k + A},
\label{eq:vanalbada}
\end{equation}
where $A$ and $B$ are defined in~\eqref{eq:AB}. $k$ should be a positive integer that satisfies
\begin{equation}
B \leq 2\Big(1+\frac{1}{k}\Big(\frac{k-1}{k}\Big)^{k-1}\Big)^{-1} \min(1,A).
\end{equation}

The Modified Van Albada limiter introduces an additional safeguard against nonphysical behavior and is defined as
\begin{equation}
\phi_{lim}(\theta) = 
\begin{cases}
 0 & \text{if } \phi < 0,\\
 \dfrac{B (\theta^k + \theta)}{\theta^k + A} & \text{if } \phi \geq 0.
\end{cases}
\label{eq:vanalbada_mod}
\end{equation}

\end{document}